\documentclass[%
 reprint,
 amsmath,amssymb,
 aps,
]{revtex4-1}

\usepackage{graphicx}
\usepackage{dcolumn}
\usepackage{bm}

\begin{document}

\preprint{APS/123-QED}

\title{Bacteria exploit torque-induced buckling instability for flagellar wrapping}

\author{Takuro Kataoka$^{1}$, Taiju Yoneda$^{1,2}$, Daisuke Nakane$^{3}$ and Hirofumi Wada$^{1}$}
 \email{hwada@fc.ritsumei.ac.jp}
\affiliation{%
 $^{1}$ Department of Physical Sciences, Ritsumeikan University, Kusatsu, Shiga 525-8577, Japan\\
 $^{2}$ Faculty of Design, Kyushu University, Fukuoka 815-8540, Japan \\
 $^{3}$ Department of Engineering Science, The University of Electro-Communications, Tokyo, Japan
}%


\date{\today}

\begin{abstract}
Recent advances in microscopy techniques has uncovered unique aspects of flagella-driven motility in bacteria.
A remarkable example is the discovery of flagellar wrapping, a phenomenon whereby a bacterium wraps its flagellum (or flagellar bundle) around its cell body and propels itself like a corkscrew, enabling locomotion in highly viscous or confined environments.
For certain bacterial species, this flagellar-wrapping mode is crucial for establishing selective symbiotic relationships with their hosts.
The transformation of a flagellum from an extended to a folded (wrapped) state is triggered by a buckling instability driven by the motor-generated torque that unwinds the helical filament. 
This study investigated this biologically inspired, novel buckling mechanism through a combination of macroscale physical experiments, numerical simulations, and scaling theory to reveal its underlying physical principles.
Excellent quantitative agreement between experiments and numerical results showed that long-range hydrodynamic interactions (HIs) are essential for accurate quantitative descriptions of the geometrically nonlinear deformation of the helical filament during wrapping.   
By systematically analyzing extensive experimental and numerical data, we constructed a stability diagram that rationalized the stability boundary through an elastohydrodynamic scaling analysis. 
Leveraging the scaling nature of this study, we compared our physical results with available biological data and demonstrated that bacteria exploit motor-induced buckling instability to initiate their flagellar wrapping.
Our findings indicate that this mechanically-driven process is essential to bacterial-wrapping motility and consequently, plays a critical role in symbiosis and infection.
\end{abstract}

\pacs{Valid PACS appear here}
\maketitle


\section{Introduction}
Flagella-driven bacterial motility~\cite{Berg-Nature-1973, Wadhwa-Review-2022} has long inspired the investigation of various exciting physical phenomena such as the low-Reynolds-number propulsion~\cite{Taylor-RS-1951, Lighthill-SIAM-1976, Purcell-AJP-1977, Lauga-Powers-2009, Lauga-Review-2016}, elasto-hydrodynamic instabilities of helices~\cite{Hotani-JMB-1982, Coombs-PRL-2002}, statistical physics of chemotaxis~\cite{Berg-Book-2004}, and fluid dynamics of bacterial suspensions~\cite{Childress-Book-1981, Dombrowski-PRL-2004, Ishikawa-Review-2009}. 
Biologically, bacteria move through different physical environments for various reasons, such as foraging for food, avoiding repellents, or establishing infection or symbiosis with their hosts~\cite{Raina-Review-2019}. 
Their movement occurs not only in open water environments, but also through complex, highly structured, viscoelastic environments and/or strongly confined spaces, as encountered in soil, mucus, and tissue~\cite{Biondi-AIChE-1998, Harman-PNAS-2012, Lauga-Review-2016, Arratia-Review-2022}. 
Recent advances in microscopy have revealed specialized strategies by which bacteria use their flagella to achieve motility in such unique physical environments~\cite{Grognot-Review-2021}. 
Among these techniques, flagellar wrapping motility is particularly remarkable~\cite{Thormann-Review-2022}. 
When the motor changes the rotational direction, the extended configuration of the helical filament becomes unstable and the filament subsequently wraps around the cell body~\cite{Kuhn-PNAS-2017, Hintsche-SciRep-2017, Kinosita-ISME-2018}. 
As the flagellum and cell body rotate in opposite directions, the bacterium reverses its translational direction, propelling itself like a screw. 
The wrapping motility has various applications. For instance, {\it Shewanella putrefaciens} applies it to escape from physical traps, and {\it Caballeronia insecticola} applies it to pass through constrictions, reaching the symbiotic organ of an insect~\cite{Ohbayashi-PNAS-2015, Kinosita-ISME-2018}. Therefore, the wrapping may have significant consequences on the flagella-mediated motility in a range of bacteria.

\begin{figure}
\centering
\includegraphics[width=.95\linewidth]{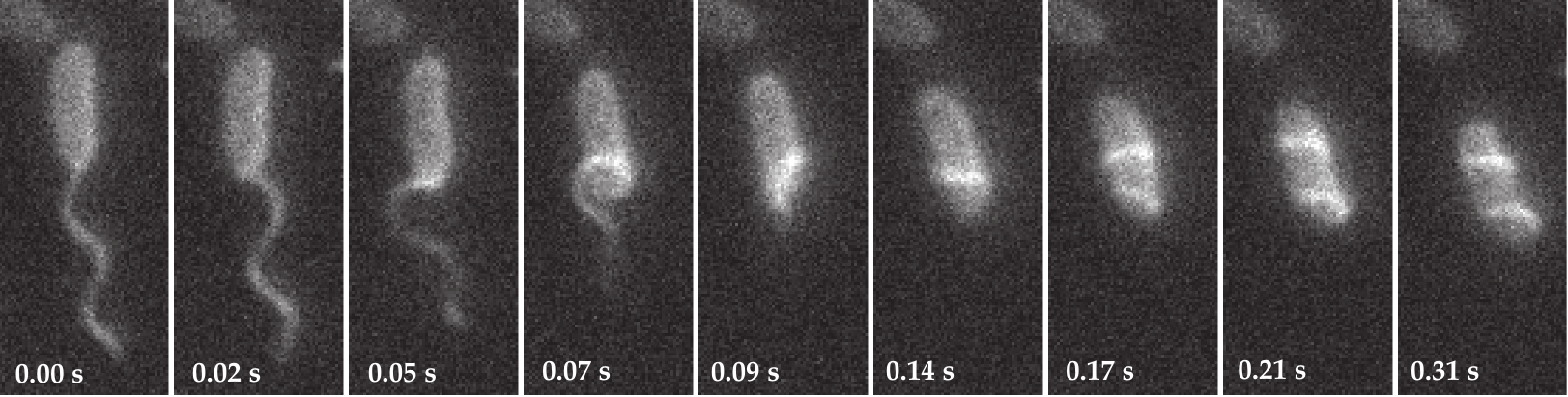}
\caption{Sequential fluorescent images of the wrapping dynamics of free-swimming {\it Caballeronia insecticola}.
When the rotary motors of the bacteria switch their rotational direction from counterclockwise to clockwise, a flagellar filament (or a bundle of them) undergoes a normal to coil polymorphic transformation (both left-handed helices). This corresponds to the leftmost image at 0.00 s.
Subsequently, buckling instability occurs in the filaments, causing them to wrap around the cell body.
This process can be completed within approximately 0.2-0.3 s. See also SI Movie 1.}
\label{fig:schematic}
\end{figure}

A typical wrapping process of {\it C. insecticola} is displayed in Fig.~\ref{fig:schematic} and SI Movie1. 
When a swimming cell stops, its motor changes its rotational direction from counterclockwise (CCW) to clockwise (CW) (viewed from the free end of the flagellum).
As elastic stresses balance viscous stresses on the rotating filaments, the flagellar filaments undergo normal-to-coil polymorphic transformation (both left-handed (LH)), in which the helical radius increases while the pitch shortens. 
The CW rotation of the motor further induces buckling instability in the filaments of the coil state, eventually causing the filaments to wrap around the cell body.  
Notably, previous numerical studies have reproduced the wrapping of helical filaments~\cite{Kuhn-PNAS-2017, Park-SciRep-2022}, and a qualitative sketch of the wrapping mechanism has also been proposed~\cite{Thormann-Review-2022}.
However, to the best of our knowledge, an accurate quantification of this biologically-inspired, novel type of helical buckling has not been performed. 

The aim of this study is to investigate this unconventional buckling by combining a macroscale physical experiment, numerical simulation, and scaling theory to reveal its physical mechanism. Considering the extreme thinness of the flagellar filament (approximately $20$ nm) and the rapid timescale of wrapping process (completes within 0.2 s, with a relatively high motor rotation rate of $\sim 50$ Hz), direct quantification of the wrapping under a microscope is challenging. 
Therefore, we constructed a table-top model system that reproduces the wrapping transition of a soft helix, enabling us to directly link the mechanics of 
this unique slender structure with the wrapping motility in bacteria.
Approaches based on scale analog models provide insight into understanding the mechanics of shape instability, propulsion, and bundling of flagellar filaments~\cite{Macnab-PNAS-1977, Kim-PNAS-2003, Liu-PNAS-2011, Rodenborn-PNAS-2013, Jawed-PRL-2015}.
Notably, Jawed {\it et al} investigated the shape instability of a flexible rotationally driven helix at its one end in a viscous fluid~\cite{Jawed-PRL-2015}.
They focused on the rotation opposite to that in our study (with respect to the helical handedness), thereby addressing the buckling instability at higher frequencies than those reported in the present study.

\begin{figure}
\centering
\includegraphics[width=.85\linewidth]{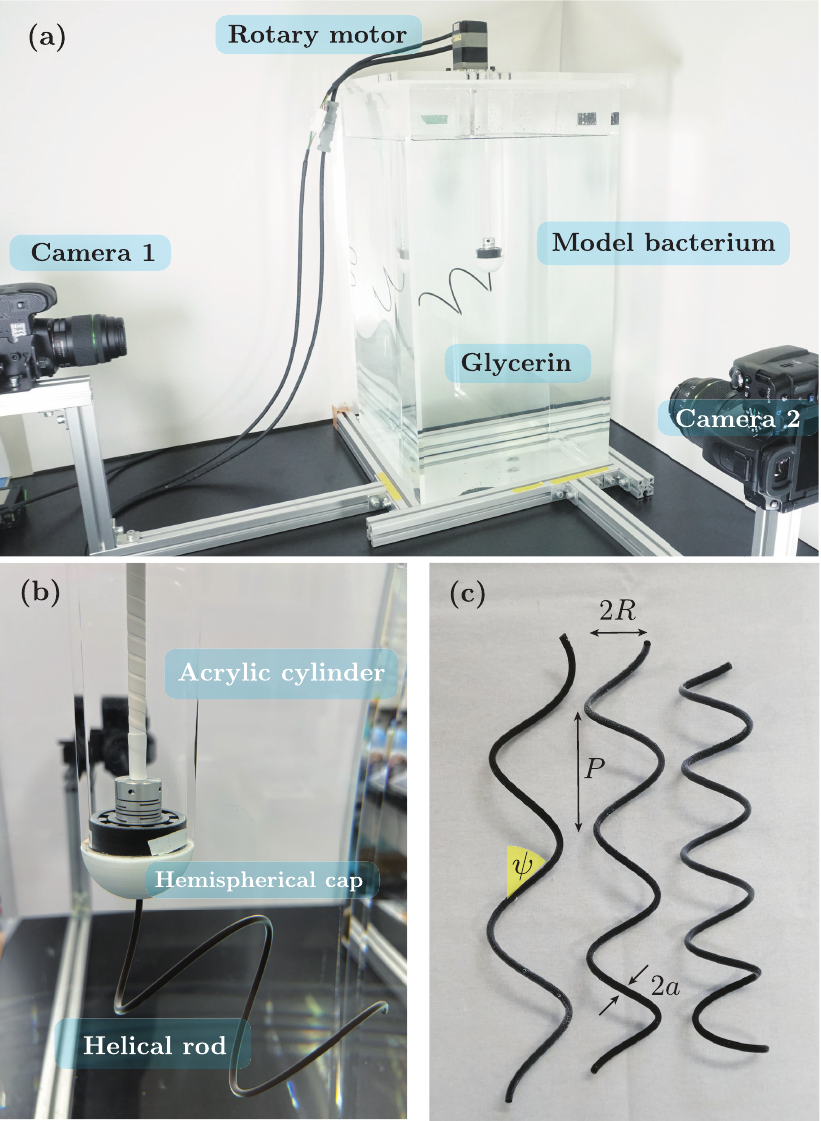}
\caption{Overview of experimental system: (a) Flexible helical filament is clamped and rotated via a rotary motor 
in a tank filled with glycerin. Two cameras positioned at right angles capture the helix configuration simultaneously. 
(b) Enlarged view of the critical parts of the experiment comprising helix, central shaft, plastic cap, and acrylic cylinder.
(c) Photographs of representative fabricated silicone-based helices of various lengths and pitch angles (all left-handed), with the definition 
of geometric parameters (helical radius $R$, pitch $P$, pitch angle $\psi$, and radius of circular cross-section $a$).}
\label{fig:experiment}
\end{figure}

Moreover, in contrast to all previous attempts, our scale model comprises a flexible helical filament and rigid cylindrical body around which the helix can wrap.
See Fig.~\ref{fig:experiment} (b).
To complement the experiment, we performed numerical simulations by combining the Kirchhoff elastic rod formulation~\cite{Audoly-Book} and Stokesian dynamics simulation method~\cite{Manghi-Review-2006}.
Thus, we could extrapolate our experimental results at a low but finite Reynolds number, $Re\sim 10^{-2}$, to a regime relevant to microbiology, $Re < 10^{-4}$.
Our analysis focused on the buckling instability that defines the initial stage of the wrapping process. 
By compiling experimental and numerical data, we established a stability diagram, which was rationalized by the elastohydrodynamic scaling theory.

A polar flagellated bacterium may have several flagella at one pole. These flagella can form a bundle during propulsion.
In this study, we considered the bundle of flagella as a single elastic helical filament. 
This simplification is biologically feasible, as a flagellar bundle of {\it C. insecticola} remains stable even at stalled rotary motor conditions~\cite{Yoshioka-preprint-2025}.
For a free-swimming bacterium, the cell body experiences counter-force and -torque such that the total force and torque on the cell vanish.
However, we ignored the effects of cell body rotation and translation because we fixed the position of the cylinder in a tank.
Although the proposed numerical simulation showed that the counter motion of the cell body could slightly increase the success rate of flagellar wrapping~\cite{Yoshioka-preprint-2025}, this subtle issue was less important for the flagellar buckling addressed in this study.

\section{Elasticity of helices}
\begin{figure}
\centering
\includegraphics[width=.90\linewidth]{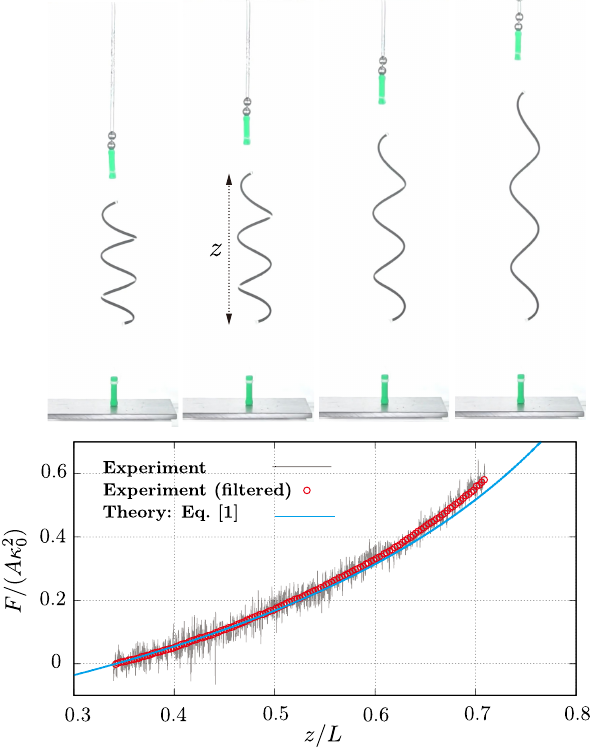}
\caption{Uniaxial stretching test of the fabricated helix. 
(a) Photographs of the helix under the increasing extension $z/L$ indicated in the figure, where $z$ represents the end-to-end distance of the helix.
(b) Measured force $F$ (rescaled by $A\kappa_0^2$) as a function of the rescaled extension $z/L$. 
The gray solid line shows the raw experimental data, and the red symbols represent the same experimental data subjected to low-pass filtering at 10 Hz.
The blue solid shows the analytical prediction expressed in Eq.~(\ref{eq:Force}), with the values of the parameters targeted in the fabrication process.}
\label{fig:exp_stretch}
\end{figure}

Various helices were custom-made in our laboratory (Fig.~\ref{fig:experiment} (c)), following the procedure described in Methods~\ref{sec:fabrication}.
To confirm their linear elasticities, we performed a uniaxial stretching test of the fabricated helices [Fig.~\ref{fig:exp_stretch} (a) and (b)].
The experiment was conducted in a glycerin bath, i.e., without gravitational force, such that both ends could freely rotate (i.e., moment-free).
The top end of a helix is pulled vertically quasi-statically with the bottom end being fixed at its position.
The resulting force curve [Fig.~\ref{fig:exp_stretch} (b)] is compared with the analytical expression obtained under the assumption of uniform deformation~\cite{Wada-EPL-2007},
\begin{equation}
 \frac{F}{A\kappa_0^2} = \Gamma\frac{(\cos\psi-\cot\psi_0\sin\psi)(\sin\psi+\Gamma\cot\psi_0\cos\psi)}
 	{\sin\psi (\sin^2\psi+\Gamma\cos^2\psi)^2}
 \label{eq:Force}
\end{equation}
with the extension $z/L = \cos\psi+F/(\pi Ea^2)$, where $A = \pi/4 Ea^4$ and $C = A/(1+\nu)$ are the bending and twisting moduli of the filament, respectively, $E$ is Young's modulus, $\nu$ is the Poisson's ratio, and $\psi_0$ is the equilibrium pitch angle.
Using the parameter values targeted in the fabrication, as well as $E$ and $\nu$ determined in the independent measurements (described in SI Appendix) in Eq.~(\ref{eq:Force}), we established an excellent agreement with the experimental data in Fig.~\ref{fig:exp_stretch} (b).
(A slight deviation at higher extensions is due to the small non-isotropy of the cross-section, which is further considered in the SI Appendix.)
Although the value of $E$ is assumed to be 1.2 times larger than that determined in the separate experiment (SI Appendix), all the other parameters are remarkably consistent with those targeted in the fabrication, confirming that the proposed helices possess the desired shape and elasticity.

\section{Torque-induced buckling and wrapping}
In the wrapping experiment, a helix was rotated in the CCW direction at a given angular velocity $\omega$ for equilibration.
After 10 CCW revolutions, rotation was switched to the CW direction with the same velocity, and the subsequent time evolution was recorded.  

By adopting the helix radius $R$ and the inverse of the motor angular velocity $\omega^{-1}$ as a representative size and time scale for the Stokes flow, we estimated the Reynolds number as $Re = \rho R^2\omega/\eta$, where $\rho$ is the mass density of the surrounding liquid.
For a bacterium undergoing wrapping, $R\approx 0.6$ $\mu$m, $\omega\approx 2\pi \times 50$ Hz, $\rho = 10^3$ kg/m$^3$, and $\eta = 10^{-3}$ Pa$\cdot$s for water, we have $Re \approx 10^{-4}$.
For the proposed scaled experimental model, as $R\approx 15$ mm, $\omega =  2\pi \times 0.01$ Hz, $\rho = 1.25\times 10^3$ kg/m$^3$ and $\eta = 1.1$ Pa$\cdot$s, we have that $Re \approx 2\times 10^{-2}$, which significantly exceeds $Re$ for bacteria. However, this value is still sufficiently small to ensure that the viscous forces dominate the inertia of the helices. 

Following Ref.~\cite{Kim-PNAS-2003}, we define the nondimensional parameter $M = \eta \omega L^4/A$~\cite{Machin-JEB-1958}, which is the product of the motor angular velocity $\omega$ and bending relaxation time $\sim \eta L^4/A$ of a straight filament of length $L$ and bending rigidity $A$. 
For bacteria in water, assuming $A = 3$ pN$\cdot$$\mu$m$^2$, $\eta = 10^{-3}$ Pa$\cdot$s, $\omega = 2\pi \times 150$ Hz, $L = 7 \mu$m, we estimate $M \approx 750$. 
The buckling instability is obtained for $M > M_c$, where the critical value $M_c$ depends only on the parameters of the helical geometry as shown below.

\begin{figure*}
\centering
\includegraphics[width=0.87\linewidth]{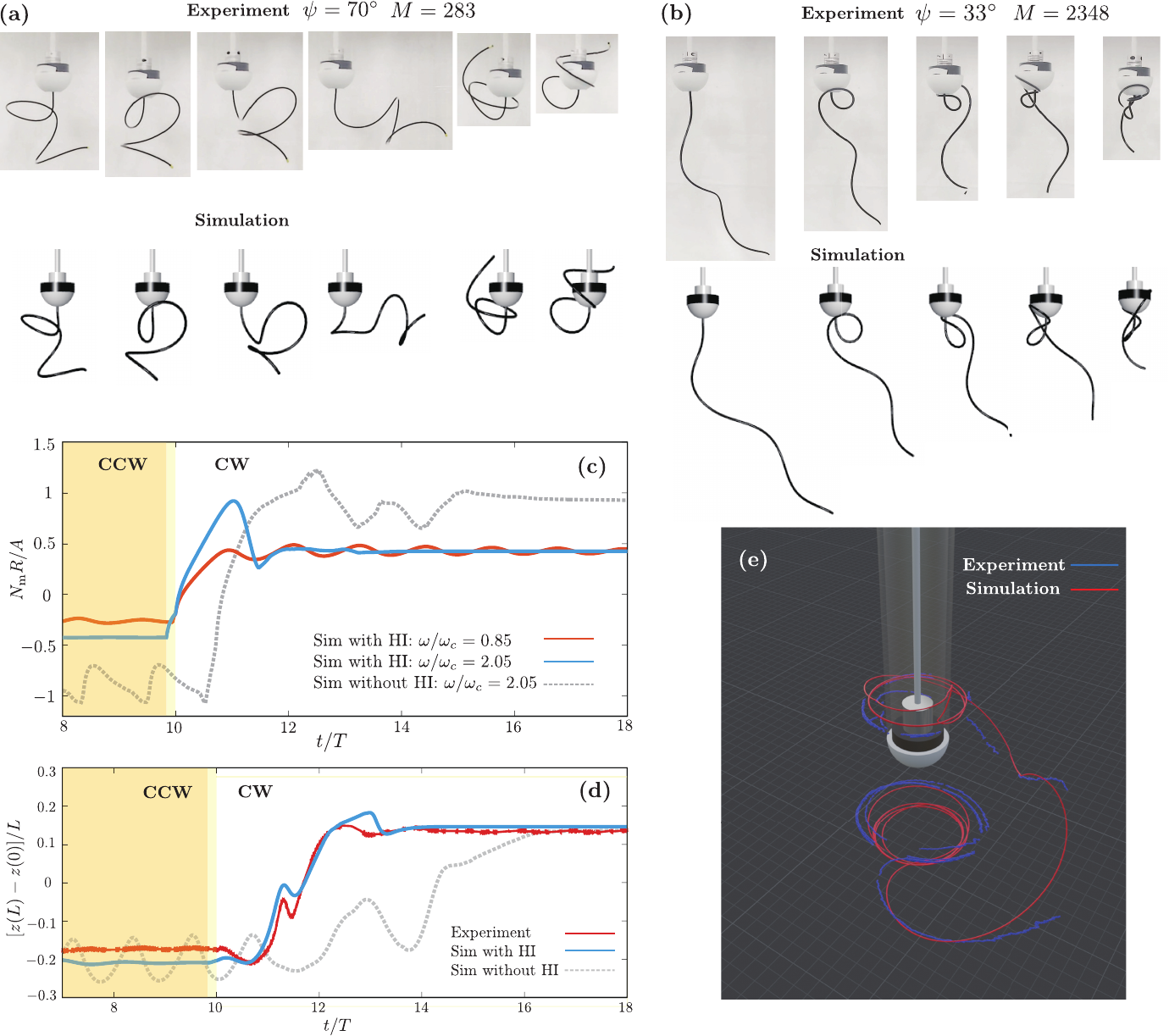}
\caption{Buckling and wrapping behavior of the rotating helix. 
(a--b) Comparison between experiment and numerical simulation (with hydrodynamic interactions, HIs) for (a) pitch angle $\psi=70^{\circ}$, helical turn $n = 2$, and $M = 283$, (b) $\psi = 33^{\circ}$, $n = 2.5$, and $M = 2348$. Time evolves from left to right. 
(c) Rescaled motor torque $N_mR/A$ that is necessary to rotate the helix at a given angular frequency $\omega/\omega_c$, for the parameter set relevant to the case in (a). 
For $\omega/\omega_c = 0.85$ (shown by the red line), the helix exhibits whirling but no buckling instability. 
For $\omega/\omega_c = 2.05$ (shown by the blue line), the helix buckles and shows wrapping as in case (a). 
The gray dashed line indicates the result from the numerical simulation without HIs for $\omega/\omega_c = 2.05$.
(d) Rescaled position of the free end of the helix as a function of rescaled time $t/T$ (or the number of revolutions of the motor), for the wrapping case shown in (a). The red line represents the experimental data, blue line represents the numerical data with HIs, and gray line represents the numerical data without HIs. (e) 3D trajectory of the tip of the filament, comparison between experiment and simulation, for case (a).}
\label{fig:snapshots}
\end{figure*}

For a small value of $\omega$, the helix maintains its LH shape, exhibiting twirling or whirling on the lower side of the tank (SI Movie 2).
By contrast, for $\omega$ exceeding a certain critical frequency $\omega_c$ (discussed subsequently), buckling occurs when the viscous force unwinds the helix considerably, with the formation of a kink connecting the LH and RH sections near the driving end, a morphology known as {\it perversion}~\cite{McMillen-JNS-2002}.
This instability leads to the wrapping of the soft helix around the cylindrical surface, as shown in Fig~\ref{fig:snapshots} (a) and SI Movie 3.
The kink is an energetically expensive localized structure; hence, the helix reconfigures itself upside down to resolve the kink.
This occurs as the tip of the helix is free. However, if both ends are constrained, the kink may rather propagate along the contour such that the RH section may prevail, as recently reported in~\cite{Dilly-PRL-2023}.
For some parameters, a helix may fail to wrap around the cylinder, ending up with the folding configuration (SI Movie 4).
The possible reasons for this are discussed subsequently.

We monitored the 3D position of the free end, ${\bf r}(L)$, and compared the experimental data (for details, see SI) with those from our numerical simulation that appropriately consider long-range hydrodynamic interactions (HIs) between the distant segments of the helix.
An excellent quantitative agreement was observed between the experiment and numerical simulation without any adjustable parameters [Fig.~\ref{fig:snapshots} (c),(d) and (e)]. 
To demonstrate the necessity of the fluid flow, we conducted a simulation using the same parameter sets while switching off the long-ranged parts of the HIs.
Data without HIs (Fig.~\ref{fig:snapshots} (d)) provide a considerably less satisfactory prediction, demonstrating that long-ranged HIs are essential for a quantitative description of the wrapping.
Although not directly related to the bacterial wrapping, the buckling of a normal-formed extended helix was tested (Fig.~\ref{fig:snapshots} (b) and SI Movie 5).
As a significantly high rotation (thus torque) is necessary for the instability, the filament is twisted close to its driving end, at which a plectoneme-like morphology appears~\cite{Wada-EPL-2009, Bruss-PRR-2019}. 
The experiment and elastohydrodynamic simulation were consistent, even in a parameter region that is extremely different from the biologically relevant regime.   

Having established the accuracy of the proposed numerical method, we investigate the motor torque $N_m$ necessary to rotate the entire helix at a given $\omega$.
As $N_m$ is extremely small to be measured experimentally, we plot the numerically computed $N_m$ in Fig.~\ref{fig:snapshots} (c) for varying $\omega$.
For $\omega/\omega_c = 0.85$, the helix is stable, exhibiting twirling motion only, for which $N_mR/A \sim 0.5$. 
By contrast, for $\omega/\omega_c = 2.05$, the helix buckles and indicates wrapping (Fig~\ref{fig:snapshots} (a)).
At the first complete CW rotation, i.e., $t/T\sim 11$, $N_m R/A$ assumes its maximum value $\sim 1$ and reaches a stationary value $\sim 0.5$ when the helix wraps around the cylinder.
If the HI is switched off (gray dashed line in Fig.~\ref{fig:snapshots} (c)), the torque $N_m$ is generally considerably large, indicating that the long-range part of HIs significantly reduces the coefficient of the effective rotational friction of a flexible helix~\cite{Wada-EPL-2009}. 
This might assist a bacterium to rapidly complete its wrapping (typically within 0.2-0.3s, Fig.~\ref{fig:schematic}).

\section{Scaling argument and diagram}
Focusing on buckling instability, we developed a scaling argument to rationalize a series of our experimental and numerical data. 
As aforementioned, the onset of the instability can be characterized by a kink between opposite handedness.
A necessary torque to generate such a kink may be $\sim A/R$, as shown in Fig~\ref{fig:snapshots} (c).
The work done by the motor per time is given by $P\sim A\omega/R$.
The viscous power dissipation of a rotating helix can be estimated as $P_v \sim \zeta R^2\omega^2 L$, where 
$\zeta = \pi\eta/\log (0.18 P/a)$ is the friction coefficient in the resistive force theory~\cite{Lighthill-SIAM-1976}.
Balancing these two values at a stationary state, we obtain the scaling prediction for the critical frequency as
\begin{equation}
 \omega_c \sim \frac{A}{\zeta R^3 L} \sim \frac{E}{\eta} \left(\frac{a}{R}\right)^3\left(\frac{a}{L}\right),
 \label{eq:scaling}
\end{equation}
where the equation $A = \pi Ea^4/4$ is used at the second equality.
The second expression indicates that in addition to the geometric parameters relying only on the aspect ratios, $R/a$ and $L/a$, the time scale of the critical frequency is considerably determined by $E/\eta$.
Using the typical values for bacteria, $E\sim 4\times10^8$ Pa, and $\eta\sim 10^{-2}-10^{-3}$ Pa$\cdot$s, as well as $R/a\sim 10^2$ and $L/a\sim 10^3$, we obtain $\omega_c\approx 10^2-10^3$ Hz, which is consistent with the observations.
In our scaled experiment, $E$ and $\eta$ are considerably different from those for bacteria. 
Plugging the measured values $E = 1.3$ MPa and $\eta = 1.1$ Pa$\cdot$s, with $R/a\sim 20$ and $L/a\sim 2\times 10^2$, we obtain $\omega_c \sim 0.01 - 0.1$Hz, which is also consistent with the experiment.
Although both time scales differ by a factor of $10^4$, they describe the same physics.  

\begin{figure*}
\centering
\includegraphics[width=0.87\linewidth]{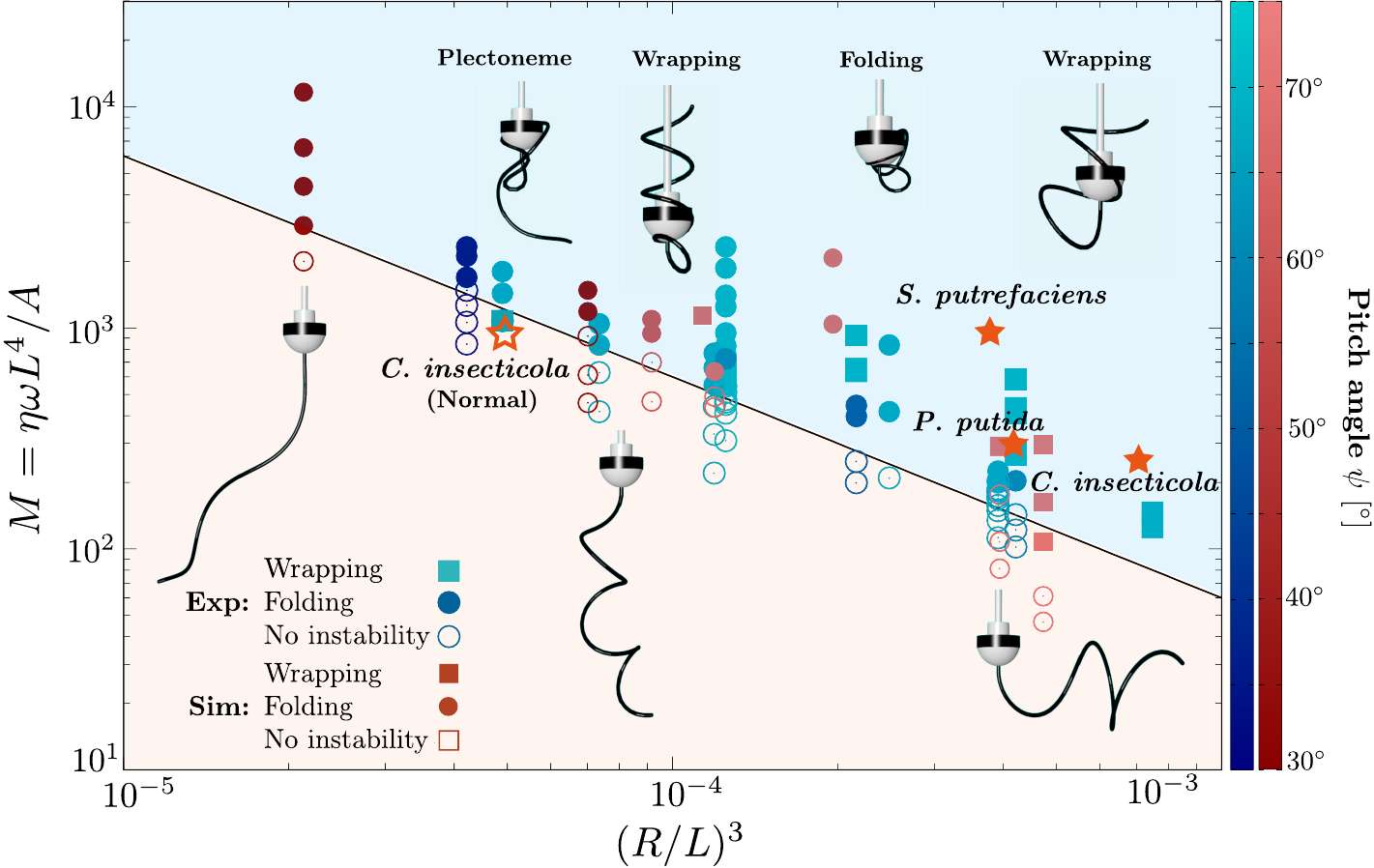}
\caption{Stability phase diagram of a helix subjected to an unwinding torque in a viscous fluid, constructed from a series of macroscale experiments and numerical simulations, on the plane spanned by the non-dimensional angular frequency $M = \eta \omega L^4/A$ and geometrical parameter of the helix $(R/L)^3$. The solid line represents the scaling prediction given by $M_c\sim (R/L)^{-3}$, with the prefactor 0.06. 
In the lower region (pale red), the left-handed helical shape is stable and undergoes a whirling motion (shown as open circle symbols).
In the upper region (pale blue), the helix undergoes buckling instability and exhibits either wrapping (filled square symbols) or folding (i.e., incomplete wrapping) (filled circle symbols). The color bar shows the magnitude of the helical pitch angle $\psi$ indicated in the colors of the symbols.
Blue and red color distinctions apply to the experimental and numerical data, respectively. Star symbols represent the data for wrapping bacteria (the details of which are given in SI). }
\label{fig:diagram}
\end{figure*}

In terms of $M$, the scaling prediction Eq.~(\ref{eq:scaling}) can be expressed as $M_c \sim (R/L)^{-3}$, which is verified subsequently.
Integrating experimental and numerical data for various sets of parameters, we constructed the stability diagram in Fig.~\ref{fig:diagram}
with the filled and open symbols representing buckling instabilities and no instabilities, respectively. 
The solid line in Fig.~\ref{fig:diagram} represents the theoretical prediction $M_c = 0.06 \times (R/L)^{-3}$, which is validated by the experimental and simulation data.
To examine the biological relevance of this study, we superpose the available bacterial data (see SI for data analysis) on the scaling plot. Results show that all three types of wrapping bacteria are located in the instability region.
Furthermore, we plot the data of the normal form of {\it C. insecticola}, which is located in the stable region, as expected. 
Summarily, bacteria can exploit the motor-driven buckling instability of a helix to realize their wrapping motility mode.

The fraction of wrapping bacterial cells increases with increasing media viscosity~\cite{Kuhn-PNAS-2017, Kinosita-ISME-2018}.
The schematic in Fig.~\ref{fig:diagram} is consistent with these observations as a large value of $\eta$ implies a large value of $M$ and the helix may become unstable for a given $\omega$. 
Fig.~\ref{fig:diagram} indicates another important aspect, that buckling occurs for a small value of $\omega$ for helices with large radii $R$.
This provides a direct mechanical explanation of the normal-to-coil polymorphic transition exhibited by all wrapping bacteria known thus far at CW motor rotation.
Considering the maximal torque of the bacterial motor, the coiled form may be essential for its wrapping.
In Fig.~\ref{fig:snapshots} (c), the critical torque for the buckling is approximately $N_c \simeq 0.5 A/R$.
Assuming $A \sim 3.0$ pN$\cdot\mu$m$^2$~\cite{Darnton-BiophysJ-2007, Shen-BiophysJ-2022} and $R \sim 0.6\,\mu$m, we have $N_c \sim 2500$ pN$\cdot$nm.
Furthermore, the peak torque for $\omega/\omega_c\approx 2$ is illustrated in Fig.~\ref{fig:snapshots} (c) as $N_m\simeq 0.9 A/R$, giving $\sim 4500$ pN$\cdot$nm.
The estimated values slightly exceed the previously reported values 1300--1800 pN$\cdot$nm of {\it Salmonella} and {\it E. coli}~\cite{Sowa-Review-2008} 
and are comparable to the values for bacteria with high-torque motors including $\sim 3600$ pN$\cdot$nm of {\it H. pylori}~\cite{Celli-PNAS-2009}, $\sim 4000$ pN$\cdot$nm of spirochete {\it Leptospira}~\cite{Nakamura-BiophysJ-2014}, and $\sim 2000-4000$ pN$\cdot$nm of {\it Vibrio} spp~\cite{Sowa-JMB-2003}.
Notably, for the normal-formed filaments of $R\sim 0.2\,\mu$m, the wrapping torque is too large to be attained by any bacterial motors. 

\section{Slow dynamics right above the onset of buckling}
\begin{figure}
\centering
\includegraphics[width=.87\linewidth]{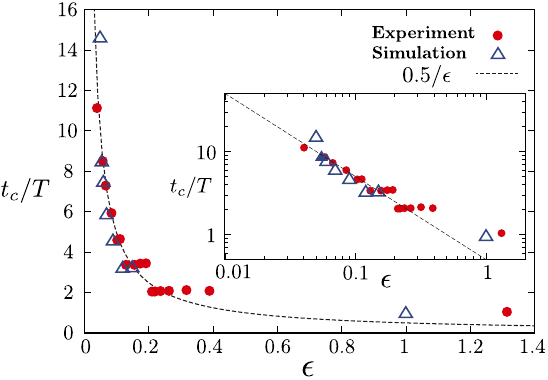}
\caption{Critical dynamics of buckling: Nondimensionalized waiting time $t_c/T$ as a function of supercritical frequency 
defined as $\epsilon = (\omega-\omega_c)/\omega_c$, for helices with a pitch angle $\psi = 70^{\circ}$. Experimental and numerical 
data are represented by filled circle (red) and open triangle (blue) symbols, respectively. The dashed line represents $0.5/\epsilon$, implying that 
the critical scaling expressed in Eq.~(\ref{eq:t_c}) is consistent with the supercritical Hopf bifurcation. Inset shows the log--log plot of the same data
highlighting the scaling of slope $-1$.}
\label{fig:waitingtime}
\end{figure}

Finally, we leverage the slow dynamics of the proposed scaled model to address the critical nature of buckling instability.
For $\omega$ slightly exceeding $\omega_c$, a considerably long waiting time is required after the CCW to CW switch until the free end of the helix starts moving upward.
However, the waiting time significantly reduces as $\omega$ increases.  
To quantify this observation, we plot in Fig.~S7 (a) $z(L)$ as a function of the number of revolutions $t/T$ from $t = 0$ at which the motor starts rotating CW, for varying $\omega (>\omega_c)$. 
In the experiment and numerical simulation, we fix the parameters of the helix shape as $n = 3$, $\psi = 70^{\circ}$.
As illustrated in Fig.~\ref{fig:waitingtime}, we plot $t_c$ at which the free end starts moving upward (indicated with the filled red symbols in Fig.~S7 (a)) as 
a function of {\it supercriticality} defined as $\epsilon = (\omega-\omega_c)/\omega_c$.
The numerical and experimental data indicate a critical slowing down consistent with the supercritical Hopf bifurcation type proposed by~\cite{Gross-Book-2009}
\begin{eqnarray}
 t_c/T &\sim& \epsilon^{-1}.
 \label{eq:t_c}
\end{eqnarray}
This observation is also consistent with those in previous analytical studies on the whirling instability of a {\it straight} elastic filament~\cite{Wolgemuth-PRL-2000, Wada-EPL-2006}.
An extension of the theories to helices is beyond the scope of the present study and will be addressed in future studies.  
Fig.~\ref{fig:waitingtime} also indicates that $t_c/T <1$ for $\omega/\omega_c>2$.
This regime is relevant to real bacteria that have to wrap their flagella rapidly. 
See Fig.~\ref{fig:diagram}.

\section{Discussion and Summary}
In this study, we investigated a novel type of dynamical buckling of a helix rotating in a viscous fluid, by combining experiments, numerical simulations, and scaling theories. 
Leveraging the scaling nature of our results, we elucidated the physical mechanism underlying the flagellar instability that defines the onset of bacterial wrapping motility.
In particular, we showed that a flagellar conformation with an increased helical radius was necessary for the filament to achieve wrapping transition.
Such a polymorphic transformation may be stimulated by a CW torque, indicating the importance of the microscopic structure and architecture of flagellar filaments. 
Further, we suggested that flagellar-wrapping bacteria should possess high-torque rotary motors for their output comparable to or exceeding 4,000 pN$\cdot$ nm.
In addition to the aforementioned mechanical requirements, the hook at the base of the flagellum is an essential mechanical component for wrapping motility.    

In the proposed scaled model, a characteristic loop was formed in proximity to the driving end because the helix end was vertically clamped.
See Fig.~\ref{fig:snapshots} (a) and Fig.~\ref{fig:diagram}.
 However, such a bending loop has not yet been observed in real bacteria.
A bacterium has a tiny mechanical joint, known as a hook, with a length of approximately $\ell_{\rm H}\sim 50$ nm that connects the filament at its base to the rotary motor~\cite{Mondino-Review-2022}. 
Owing to its flexibility, the hook functions as a universal joint, transmitting the motor torque to the filament even when the filament is highly bent away from the rotational axis of the motor.
Moreover, it accounts for the "flicking"~\cite{Vogel-EPJE-2012, Son-NatPhys-2013, Jabbarzadeh-PRE-2018} and complex surface motility~\cite{Anda-ACS-2017, Ishimoto-JFM-2019} of polar bacteria.
Thus, the hook enables a bacterium to wrap its filament around its cell body without forming any localized bend. 
Our clamped end differs from a bacterial hook, imposing a qualitatively different boundary condition for the helix dynamics.
This difference may account for the folding configuration (or incomplete wrapping) in our scale model, which is rarely observed in actual bacteria. 

Considering a hook/filament length ratio of typically $\ell_{\rm H}/L \sim 10^{-2}$, the joint should be virtually length-less, yet fully torque transmittable, spherically rotatable, and mechanically tolerant.
Although such a multi-functional mechanical joint cannot easily be developed, an engineered hook-like joint can be a key mechanical component for developing future artificial machines and micro robots~\cite{Huang-SciAdv-2019} that can wrap their flagellum-like appendices to move in highly viscous or confined environments.  

Another major concern is determining the amount of propulsive force generated by the wrapping mode. 
Given that several symbiotic bacteria such as {\it Vibrio fischeri} and {\it Caballeronia insecticola} adopt a wrapping motility mode to achieve directed movement in narrow passages in host organs for establishing a symbiotic relationship, the effects of spatial confinement on swimming speed and efficiency should be investigated~\cite{Gidituri-JFM-2024}.
A spatial confinement will be relatively easily implemented. However, visualizing a flow field and/or measuring tiny propulsive forces generated in the wrapping mode will be challenging.
Numerically, the Stokesian dynamic approach would be inapplicable to problems involving no-slip boundaries on complex geometry.
Alternative approaches, such as the boundary element method, which is currently in progress, will be suitable~\cite{Yoshioka-preprint-2025}.

\section{Materials and Methods}
\subsection{Macroscale model}
In an acrylic tank of dimensions 300 mm $\times$ 300 mm $\times$ 600 mm, we built a model bacterium comprising a rigid cell body and flexible elastomeric helix, which were immersed in approximately 50 L liquid glycerin. 
A rigid acrylic hollow cylinder of radius 5 cm and length 25 cm was vertically immersed in glycerin at the center of the tank, whose upper end was attached to a covering plate over the tank. The lower end was covered with a 3D-printed hemispherical cap of radius 5 cm. 
This rigid cylinder with the capped hemispherical end mimics the bacterial cell body around which a helical filament can wrap. 

\subsection{Fabrication of helices}~\label{sec:fabrication}
In our laboratory, a uniform helical filament was custom-made from elastomer HTV-4000 (Young's modulus 0.9 MPa and density 1.15 g/cm$^3$). 
First, a plastic tube was wrapped around a 3D-printed cylindrical object with a helical groove on its surface, preventing the flattening of the cross-section of the tube. 
Then, the wrapped object was used as a mold to form a helix. 
To match the density of the helix with that of glycerin, we followed the procedure proposed in Ref.~\cite{Jawed-PRL-2015} and added iron oxide powder (density 5.74 g/cm$^3$) to the HTV-4000 polymer before mixing it with the base for polymerization~\cite{Sano-JPMS-2022}. 
The resulting density mismatch was typically 1--2\% (and always less than 5\%), minimizing the effects of gravity or buoyancy.
We fabricated various helices with total length $L = 200-500$ mm, pitch angle $\psi = 33^{\circ}-70^{\circ}$, helical radius $R = 12.5-32.5$ mm, and pitch $P = 34-120$ mm, with an isotropic cross-sectional radius $a = 1.5$ mm.
A helical geometry corresponding to a coil-formed flagellum with a pitch angle $\psi = 69-70^{\circ}$ and 2 or 3 full turns, was mainly investigated. However, other helical forms were also investigated. 

\subsection{Rotating experiment}
A helix was clamped at one end to the bottom of the hemispherical cap through a rotational coupling.
A torque generated by a stepping motor mounted just upon the top cover of the tank was transmitted via a long shaft running vertically inside the hollow cylinder to the helix. 
From a series of independent measurements, Young's modulus of our iron-filled elastomer was $E = 1.5$ MPa, the density of the glycerin was $\rho_{\rm gly}$ = 1.25 g/cm$^3$, and the shear viscosity of the glycerin was $\eta = 1.1$ Pa$\cdot$s at room temperature $19^{\circ}$, in agreement with the literature.
We conducted our experiment in an air-conditioned room (at a temperature typically in the range $19^{\circ}-23^{\circ}$ C), without controlling the temperature of the system.
The temperature dependence of the shear viscosity of glycerin was considered in the data analysis, as detailed in SI, whenever the room temperature data were available. 
Despite the temperature sensitivity of the material parameters of glycerin, the temperature variations due to viscous heating during the experiment were negligible because relevant rotational frequencies were always sufficiently low (typically less than 0.02 Hz).
During the experiment, the shape evolution of the helix was recorded using digital imaging, which was then analyzed by capturing the images from the videos.

\subsection{Stretching test}
The uniaxial stretching test of a flexible helix was performed in a smaller tank than that for the rotating experiment filled with glycerin. Thus, the effects of buoyancy or gravity on helix deformation were negligible.  
To ensure moment-free boundary conditions, both ends of the helix were tied with thin threads (of negligible twist moduli) to short rigid magnetic rods that were positionally fixed through freely rotatable magnetic beads. 
Starting from the equilibrium configuration, the top end of the helix was pulled upward at a sufficiently slow speed of $0.1$ mm/s to ensure quasi-statistical loading.

\subsection{Numerical simulation}
A helical filament was discretized into $N$ spheres of radii $a$ connected linearly via sufficiently 
stiff springs, which maintained the filament virtually inextensible.
The internal elastic force, ${\bf f}_i$, and torque about the local tangent, $m_i$, acting on the $i$-th sphere comprised bending, twisting, and stretching contributions and were calculated from the Kirchhoff elastic energy using previous variational methods established by Refs.~\cite{Chirico-Biopolymers-1994, Chirico-Biopolymers-1996, Wada-PRE-2009}.
The external force, ${\bf p}_i$, represented a repulsive force from the rigid cylindrical body (plus a small buoyancy or gravitational force). It was computed from properly defined potential energies.
In the Stokes regime, inertia is irrelevant and the filament dynamics is described by the viscous force and torque balance defined by~\cite{Manghi-Review-2006}
\begin{equation}
 {\bf v}_i = \sum_{i=1}^{N} {\boldsymbol{\mu}}_{ij}({\bf r}_{ij}) \cdot ({\bf f}_j+{\bf p}_j),
 \quad
 \omega_i = \mu_r  m_i
\end{equation}
where ${\bf v}_i = d{\bf r}_i/dt$ and $\omega_i$ are the linear and angular velocities of the $i$-th sphere, respectively, and 
${\bf r}_{ij} = {\bf r}_i-{\bf r}_j$.
Long-range HIs between two distant spheres $i$ and $j$ are included by applying the Rotne--Prager mobility tensor as
\begin{equation}
 {\boldsymbol{\mu}}_{ij} ({\bf r}) = \frac{1}{8\pi \eta r }\left[{\bf 1}+\frac{{\bf r}}{r} \frac{{\bf r}}{r} +\frac{2a^2}{r^2}
 \left(\frac{\bf 1}{3}-\frac{\bf r}{r}\frac{\bf r}{r}\right)\right],
\end{equation}
where $\eta$ is the shear viscosity of the surrounding fluid.
For overlapping spheres, i.e., $r_{ij} < a$, an appropriately modified expression is used for ${\boldsymbol{\mu}}_{ij}$.
For simplicity, the viscous rotational dynamics are assumed to be entirely local. 
The translational and rotational self-mobilities of a sphere of radius $a$ can be expressed as
${\boldsymbol{\mu}}_{ii}=(1/6\pi \eta a) {\bf 1}$ and $\mu_r= 1/8\pi \eta a^3$, respectively.
We neglected the non-slip condition imposed at the cylindrical surface, implying that the fluid flow passed freely through the cell body.
Although the hydrodynamic screening performed by the cell surface may be essential for the wrapping configuration,
the buckling behavior is unaffected by this simplification, as demonstrated by a quantitative agreement with the experiment (Fig.~\ref{fig:snapshots}).

\begin{acknowledgments}
This study was supported by the Japan Society for the Promotion of Science Grant-in-Aid for Scientific Research (JSPS KAKENHI Grants No. 22H05067 and No. 23K22463 to H.W., and Grant 22H05066 to D.N.) and JST SPRING (Grant No. JPMJSP2101to T. K.).
\end{acknowledgments}


%

\onecolumngrid
\appendix

\section{Macroscale Experiment}

\begin{figure*}[htbp]
\begin{center}
 \includegraphics[width=0.87\linewidth]{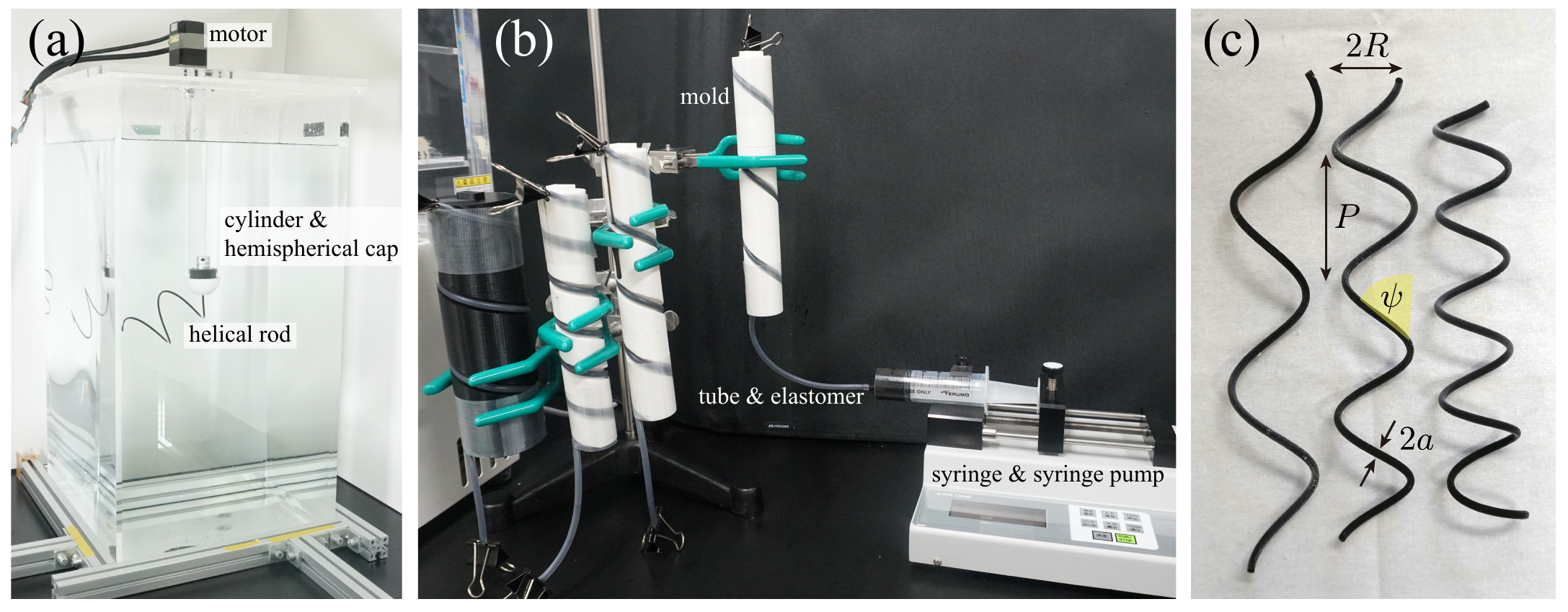}
\caption{Overview of our experimental system:
		(a) Photograph of the experimental setup.
		(b) Fabrication apparatus of a flexible helical rod.
		(c) Photographs of representative helical rods of various length and pitch angles, 
		with the definition of geometric parameters (helical radius $R$, pitch $P$, pitch angle $\psi$, and radius of the circular cross section $a$.}
\label{exp:fig:exp_setup}
\end{center}
\end{figure*}

\subsection{Experimental apparatus}
A cuboid acrylic water tank with thickness of 1 cm and inner diameter of 30 cm $\times$ 30 cm $\times$ 60 cm was filled with approximately 50 L of liquid glycerin of the mass density $\rho_{\mathrm{gly}}=1.25~\mathrm{g/cm^3}$ at temperature $T=20^{\circ}$.
An acrylic cylinder with a thickness of 4 mm, outer diameter of 5 cm, and axial length of 25 cm was attached to the center of the lid of the tank.
A hemispherical plastic cap of 4 mm thickness and 5 cm diameter was fabricated using a 3D printer (Raise 3D E2, PLA), which was then attached to cap the lower side of the cylinder.
A stepping motor (Oriental motor, Japan) was mounted on the tank lid, whose torque was transmitted via a straight shaft to a flexible helical rod (see the next section) that was suspended from the bottom of the cylinder.

\subsection{Fabrication of the helical rod}
In this study, we used the elastomer HTV-4000 (Young's modulus $E_{\mathrm{HTV}}=0.9$ MPa and mass density $\rho_{\mathrm{HTV}}=1.15~\mathrm{g/cm^3}$) to create a flexible helical rod as a model for a bacterial flagellar filament.
To create a mold for a helical rod of isotropic cross section, we first wrapped a plastic tube around a 3D-printed cylindrical object with a helical groove on its surface, which prevented flattening of the cross-section of the tube [Fig.~\ref{exp:fig:exp_setup} (b)]. 
For actual bacteria in water, the effects of gravity are negligible because of their minute size typically $\sim 1~\mu$m.
To minimize any gravitational or buoyancy effects in our scaled model, we followed the procedure described in Ref.~\cite{Jawed-PRL-2015} and added iron oxide powders ($\rho_{\mathrm{FeO}}=5.74$) to the HTV-4000 polymer before mixing it with the base for polymerization.
We mixed the base and catalyst at 1:1 for addition polymerization. 
The mixing process consisted of agitation at 2000 rpm for 40 s and centrifugal defoaming at 2200 rpm for 40 s in defoaming mode (A Thinky ARE-310, Japan).
Subsequently, the liquid was poured into a tube wrapped around the mold at 200 mL/h using a syringe pump (ASONE, SPDC-1).
Theoretically, the density of the final elastomer is determined 
by 
\begin{eqnarray}
 \rho_{\mathrm{eff}} &=& \frac{(m_{\mathrm{base}}+m_{\mathrm{cat}}+m_{\mathrm{FeO}})}{(m_{\mathrm{base}}+m_{\mathrm{cat}})/\rho_{\mathrm{HTV}} + m_{\mathrm{FeO}}/\rho_{\mathrm{FeO}}}.
\end{eqnarray}
After curing, the density of the actual iron-oxide mixed elastomer was measured, and the final density mismatch $|\rho_{\mathrm{eff}}-\rho_{\mathrm{gly}}|/\rho_{\mathrm{gly}}$ was determined as typically $1\%-2\%$ and always less than $5\%$.

As shown in Fig.~\ref{exp:fig:exp_setup} (c), we fabricated various helical rods with a total length $L=200-500$ mm, pitch angle $\psi=33^{\circ}-70^{\circ}$, helical radius $R=12.5-32.5$ mm, and pitch $P=34-120$ mm, with an isotropic cross sectional radius $a=1.5$ mm.

\section{Measurement of Young's modulus and Poisson's ratio}

\begin{figure}[htbp]
\begin{center}
 \includegraphics[width=0.999\linewidth]{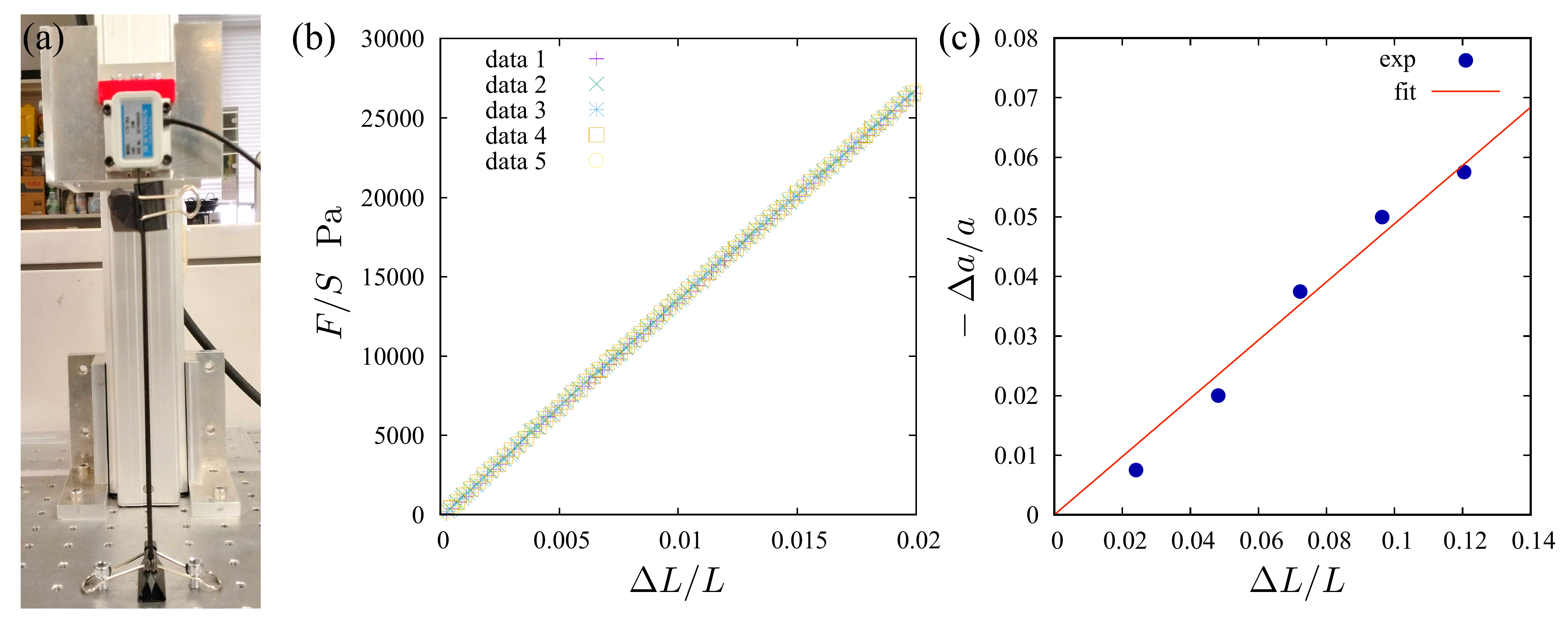}
    \caption{
		Measurement of the elastic parameters of our iron-mixed silicone elastomer: 
		(a) Photograph of tensile test system.
		(b) Stress vs. strain relationship obtained in the uniaxial stretching experiment of a rod with radius $a=1.5$ mm and total length $L=175$ mm.
		(c) Change in radius $-\Delta a/a$ of a rod cross-section as a function of the axial strain $\Delta L/L$
		measured for a rod with $a=2.0$ mm and $L=166$ mm.
	}
    \label{exp:fig_Young}
\end{center}
\end{figure}

We measured the Young's modulus $E$ and Poisson's ratio $\nu$ of the iron-oxide mixed HTV-4000 elastomer in a standard tensile test [Fig.~\ref{exp:fig_Young} (a)].
For the measurement of $E$, one end of a rod with $a=1.5$ mm and $L=175$ mm was clamped vertically and the other end was
pulled at a constant speed of $0.1~\mathrm{mm/s}$ using a stepping motor (EASM4XE040ARAC, Oriental motor),
and the tensile force $F$ was measured using a load cell (LTS-2KA, Kyowa) connected to the pulling end of the rod.
In Fig.~\ref{exp:fig_Young} (b), we plot the measured stress $F/S$, where $S=\pi a^2$ is the area of the cross section, as a function of the applied strain $\Delta L/L$, from which we determined $E=1.33\pm0.01$ MPa.

For the measurement of $\nu$, we conducted a stretching test of a rod with $a=2.0$ mm and $L=166$ mm 
and measured the change in the diameter of the cross-section $2\Delta a$ as a function of the applied strain $\Delta L/L$ [Fig.~\ref{exp:fig_Young} (c)], from which we determined $\nu=0.49$, as valid for rubber-like (incompressible) materials.

\section{Measurement of shear viscosity of glycerin}

\subsection{Sphere-dropping method}~\label{sec:method1}

\begin{figure}[htbp]
\begin{center}
 \includegraphics[width=0.999\linewidth]{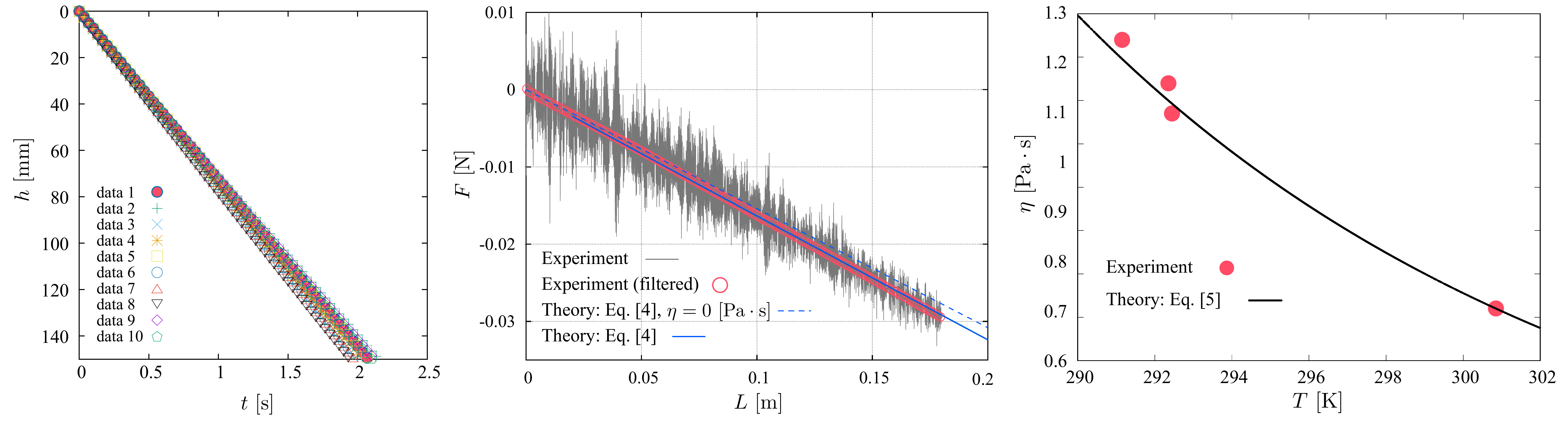}
    \caption{
		Viscosity of glycerin and its temperature dependence:
		Traveled distance of a metallic ball of radius $a=2.5$ mm and mass density $\rho=7.42~\mathrm{g/cm^3}$ falling in glycerin at 19.2 ${}^\circ$C.
		(b) Measured force $F$ acting on a slender rod of cross-sectional radius of $a=2$ mm moving in the glycerin as a function of the length $L$.
		(c) Shear viscosity of glycerin $\eta$ measured as a function of $T$. The solid line represents the fitted curve to the data based on Eq.~(\ref{eq:AndradeViscosityEquation}).} 
    \label{exp:fig:Vis}
\end{center}
\end{figure}

The shear viscosity $\eta$ of glycerin at $T=19.2^\circ$ C was determined by measuring the terminal velocity of a free-falling small sphere of mass density $\rho=7.42~\mathrm{g/cm^3}$ and radius $a=2.5$ mm.
At this low Reynolds number ($Re=\rho U_{\infty}(2a)/\eta\approx 0.3$, see below), the sphere translates at a terminal velocity $U_{\infty}$ and the Stokes drag force and gravitational force balance, which predicts
\begin{eqnarray}
	\eta &=& \frac{2}{9}\frac{(\rho-\rho_{\rm gly}) a^2 g}{U_{\infty}}.
	\label{eq:etaStokes}
\end{eqnarray}
Figure~\ref{exp:fig:Vis} (a) shows the distance vs. time of a free-falling sphere in glycerin, from which we determined the terminal velocity as $U_{\infty}=73.7\pm0.8~\mathrm{mm/s}$.
Thus, from Eq.~(\ref{eq:etaStokes}), we obtain $\eta=1.14\pm0.01~\mathrm{Pa\cdot s}$.

\subsection{Measurement of a shape-dependent friction}~\label{sec:method2}
According to the resistive force theory, when a straight rod of radius $a$ and length $L$ moves along its long axis at a constant velocity $U$, 
the viscous resistance $F_{\parallel}$ acting on the rod from a fluid of viscosity $\eta$ is given by~\cite{Lauga-Powers-2009}
\begin{eqnarray}
	F_{\parallel} &=& \frac{2\pi\eta L}{\ln{(L/a)}-1/2}U.
	\label{eq:Force:Stokes:LongRod}
\end{eqnarray}
In a 15 cm $\times$ 15 cm $\times$ 40 cm acrylic tank filled with glycerin, a rigid straight rod of SUJ2 (density $\rho=7.83~\mathrm{g/cm^3}$) with radius $a=2.0$ mm was inserted vertically into the glycerin about 1 cm away from the liquid--air interface at a constant speed $U=5$ mm/s.
In this setup, the vertical component of the total force, $F$, acting on the rod from the glycerin is the sum of viscous and buoyancy forces is
\begin{eqnarray}
	F &=&  - \rho_{\mathrm{gly}}g S L - \frac{2\pi\eta L}{\ln{(L/a)}-1/2}U,
	\label{eq:Force:Stokes:LongRod:all}
\end{eqnarray}
where $L$ is the length of the rod immersed in the glycerin. 
Note that $F$ is measured relative to the force balancing to its own weight in air, $\rho g S L_0$ (with $L_0=250$ mm the total length). 
In Fig.~\ref{exp:fig:Vis} (b), we compare Eq.~(\ref{eq:Force:Stokes:LongRod:all}) with our experimental data, and observe 
good agreement between the two for $\eta = 1.02 ~\mathrm{Pa\cdot s}$ at 19.3$^{\circ}$ C.
The result agrees well with our value for $\eta$ based on the Stokes drag formula (see the previous subsection), also confirming the validity of using Eq.~(\ref{eq:Force:Stokes:LongRod}) in the analysis of our macroscale experiment.

\subsection{Temperature dependence of viscosity}
Figure~\ref{exp:fig:Vis} (c) shows the results of the viscosity measurements at various temperatures.
The viscosity at 292 K was measured using the sphere-dropping method (see Sec.~\ref{sec:method1}), and the data at other temperatures were obtained using the force measurement of a slender rod moving in a glycerin (see Sec.~\ref{sec:method2}).

Generally, the temperature dependence of the shear viscosity of liquids may be well described by Andrade's equation~\cite{andrade1930viscosity}:
\begin{eqnarray} 
	\eta(T) &=& \eta_{\infty}\,  e^{U/k_BT},
	\label{eq:AndradeViscosityEquation}
\end{eqnarray}
where $k_B$ is the Boltzmann constant, $\eta_{\infty}$ may represent the viscosity at high enough temperature, and $U$ is the material dependent parameter of the dimension of energy. 
For our data shown in Fig.~\ref{exp:fig:Vis} (c), we find
\begin{equation}
 \eta_{\infty} = 6.92687\times 10^{-10}~\mathrm{Pa\cdot s},
 \ 
 U = 8.559\times 10^{-20}\ \mathrm{J}.
\end{equation}
Regarding our experiment described in the main text, we recorded the room temperature $T$ for all the data obtained since February 2, 2024. 
Because the glycerin was at thermal equilibrium with the ambient temperature, we used Eq.~(\ref{eq:AndradeViscosityEquation}) 
to estimate $\eta$ for the rescaling of the experimental data.
In the period from October 9, 2023 to January 31, 2024, the room temperature was unavailable.
However, because the air conditioning was always operating to maintain the ambient temperature $T$ in the range 18--20 ${}^\circ$C, we assumed 20 ${}^\circ$C as the representative value to rescale the rest of our experimental data.

\section{Detailed analysis of the uniaxial stretching test}
In the tensile test of our fabricated helical rod described in the main text, we found that the cross-section of the rod used was less isotropic than those used for the rotating experiment. 
Here, we develop a simple analytical theory of an elastic rod with an anisotropic cross-section and show an even better agreement with our experimental data to 
justify the relevant parameter values that we have estimated in the main text.

\begin{figure}[htbp]
\begin{center}
 \includegraphics[width=0.95\linewidth]{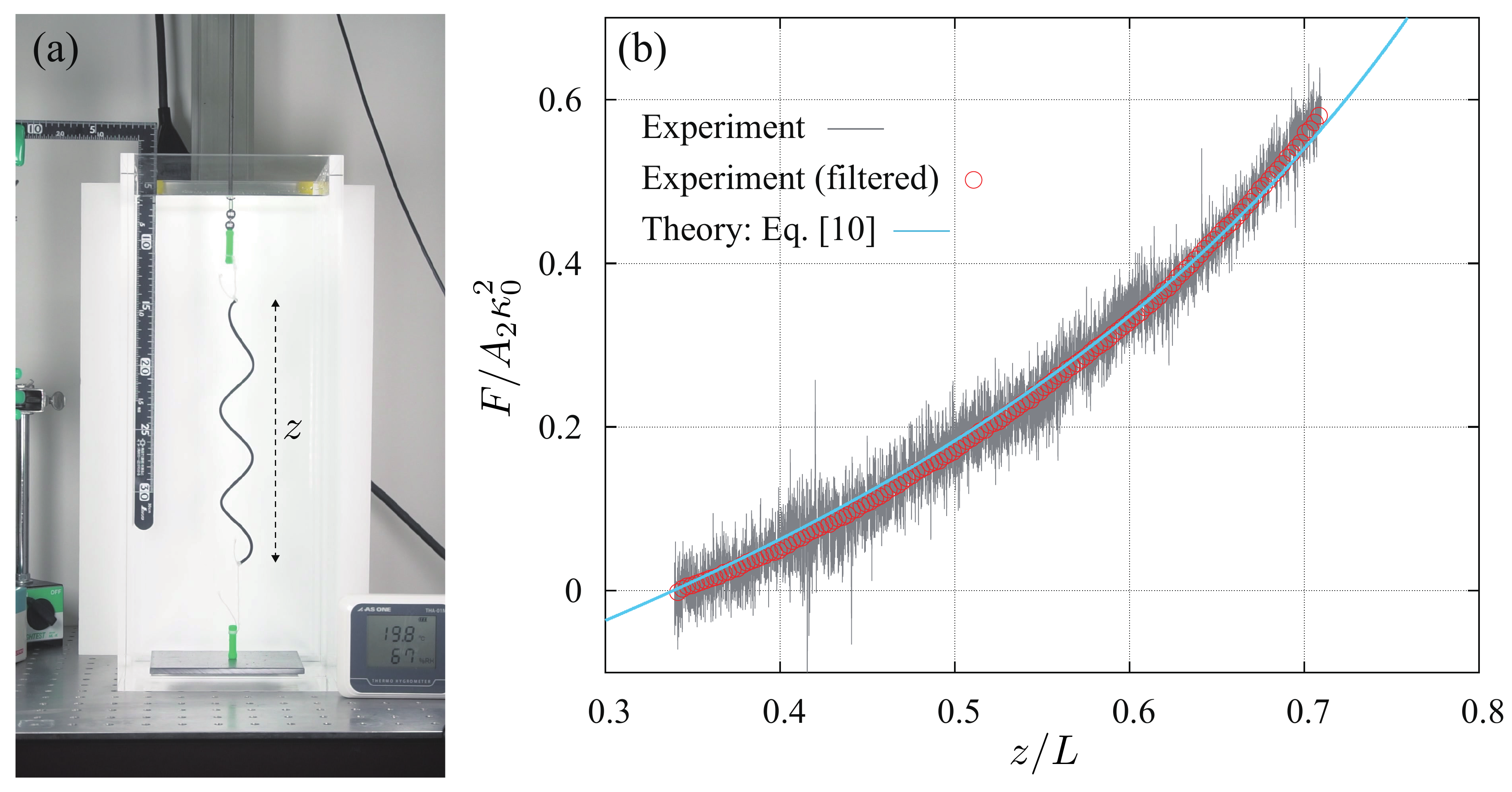}
    \caption{
		Uniaxial stretching test of a fabricated helical rod: 
		(a) Photograph of the experimental system. A helix was stretched at a constant speed of 0.1 mm/s. 
		The parameters of the helix were $2R=30$ mm, $P=34$ mm, $L=305$ mm, and $\psi=70^{\circ}$. 
		(b) Measured force F (rescaled by $A_2\kappa_0^2$) as a function of the rescaled extension $z/L$. 
		The gray solid line shows the raw experimental
		data, whereas the red symbols represent the same experimental data subjected
		to the low-pass filtering at 10 Hz. The blue solid line shows the analytical prediction given
		in Eq.~(\ref{eq:HelicalForce}), with the parameter values set in the fabrication process (except the Young's modulus $E$ set 1.2 times larger than that determined independently.)
	}
    \label{fig:fig:HelicalForce}
\end{center}
\end{figure}

Let $F$ be the external force applied at both ends of a rod of the cross section with the major radius $2a$ and minor radius $2b (<2a)$ under the torque-free boundary conditions at both ends.
The elastic deformation energy of a stretched helix of total arclength $L$ can be expressed as 
\begin{equation}
	E =\int_{0}^{L}\! \left[\frac{A_1}{2}(\kappa_1)^2+\frac{A_2}{2}(\kappa_2-\kappa_0)^2+\frac{C}{2}(\tau-\tau_0)^2 - F\cos{\psi}\right] ds.
	\label{eq:HelicalEnergy}
\end{equation}
where $s$ is the arclength of the rod centerline measured from one end of the rod, $\kappa_{1,2}(s)$ are the actual curvatures of the centerline of the rod along the two principle directions in the cross-section, $\tau(s)$ is the twist, and $\psi(s)$ is the local pitch angle. 
The two parameters, $\kappa_0=4\pi^2R/(P^2+4\pi^2R^2)$ and $\kappa_0=2\pi P/(P^2+4\pi^2R^2)$, are the spontaneous curvature and torsion, which define the stress-free helical configuration of helical radius $R$ and pitch $P$.
The bending rigidities, $A_1$ and $A_2$, and the twisting stiffness $C$ can be given by 
\begin{equation}
 A_1 = \frac{\pi}{4}Ea^3b,
 \
 A_2= \frac{3\pi}{4} Eab^3,
 \
  C = \frac{\pi}{2(1+\nu)} \frac{Ea^3b^3}{a^2+b^2}.
\end{equation}
For a uniform deformation, we have $\kappa_1=\sin^2\psi\cos\theta/R$, $\kappa_2=\sin^2\psi\sin\theta/R$ and $\tau=d\theta/ds+\sin\psi\cos\psi/R$, 
where $\theta(s)$ is the auxiliary variable.
Taking the variations in the functional $E$ with respect to $\psi$ and $\theta$, and solving the Euler--Lagrange equations, we obtain the force $(F)$ vs. extension $(z)$ relationship given by 
\begin{eqnarray} 
	\frac{z}{L} &=& \cos{\psi}+\frac{F}{\pi Eab},
	\label{eq:PitchAngleForce} \\
	\frac{F}{A_2\kappa_0^2} &=& \Gamma^{*} \frac{  ( \sin{\psi} - \lambda \cos{\psi} )  ( \cos{\psi} + \Gamma^{*} \lambda \sin{\psi}) }{ \sin{\psi}  ( \sin^2{\psi} + \Gamma^{*} \cos^2{\psi}) },
	\label{eq:HelicalForce}
\end{eqnarray}
where we have defined $\lambda=\tau_0/\kappa_0$ and $\Gamma^{*}=C/A_2$.
Note that we have added the term $F/(\pi E ab)$ to the r.h.s in Eq.~(\ref{eq:PitchAngleForce}) to consider weak axial stretching of the rod centerline. 
Using $\Gamma=C/A=1/(1+\nu)$ defined  in the main text for an isotropic rod, we have
\begin{equation} 
	\Gamma^{*}=\frac{2\Gamma}{1+(b/a)^2}.
	\label{eq:Gamma}
\end{equation}

\subsection{Uniaxial stretching test of a helix}
A 15 cm $\times$ 15 cm $\times$ 0.5 cm steel plate was placed at the bottom of a 15 cm $\times$ 15 cm $\times$ 40 cm water tank and immersed in glycerin [Fig.~\ref{fig:fig:HelicalForce} (a)].
We tied the both ends of the helix with thin threads (of negligible twist moduli) to short rigid magnetic rods.
At the bottom end, the magnetic rod stuck to the steel plate.
In contrast, the magnetic rod at the top end was connected to a long steel rod via two steel spheres, which ensured the freely rotating condition for the top end of the helix.
Finally, the long steel rod was attached to a load cell that moved vertically at a constant speed of $0.1$ mm/s driven by a stepping motor. 

The experimental force curve is shown in Fig.~\ref{fig:fig:HelicalForce} (b).
The raw data is indicated by a gray line, whereas the FFT-filtered data (at 10 Hz) are shown by red symbols, which are compared with our theoretical prediction, Eq.~(\ref{eq:HelicalForce}).
The parameter values for Eq.~(\ref{eq:HelicalForce}) are given by $\nu=0.49\pm 0.02$, $2a=3.19\pm 0.001~$ mm, $2b=2.88\pm 0.001~$ mm, $2R=30~$ mm, $P=34~$ mm, $L=305~$ mm, 
$\Gamma^{*}=0.74$, $\lambda=0.36$, and $A_2\kappa_0^2=0.020~$ N.
These are all the pre-determined or targeted values, except the value of the Young's modulus that is 1.2 times larger than the value that we determined separately. 
(This modification only influences the value of $A_2/\kappa_0^2$ in this comparison.)
The agreement between theory and experiment is excellent throughout the range studied, particularly improving the agreement at a high force regime over that shown in the main text. 
This confirms the high precision of our fabrication of the physical models of helical flagellar filaments.

\section{3D tracking using 2-way imaging}
Here, we describe the details of our stereoscopic reconstruction method of a tip of a helical rod.
The typical trajectory of the free end of the helix during the wrapping transition is shown in Fig.~4 (e) in the main text and in Fig.~\ref{exp:fig_tracking} (b).

\begin{figure*}[htbp]
\begin{center}
  \includegraphics[width=0.95\linewidth]{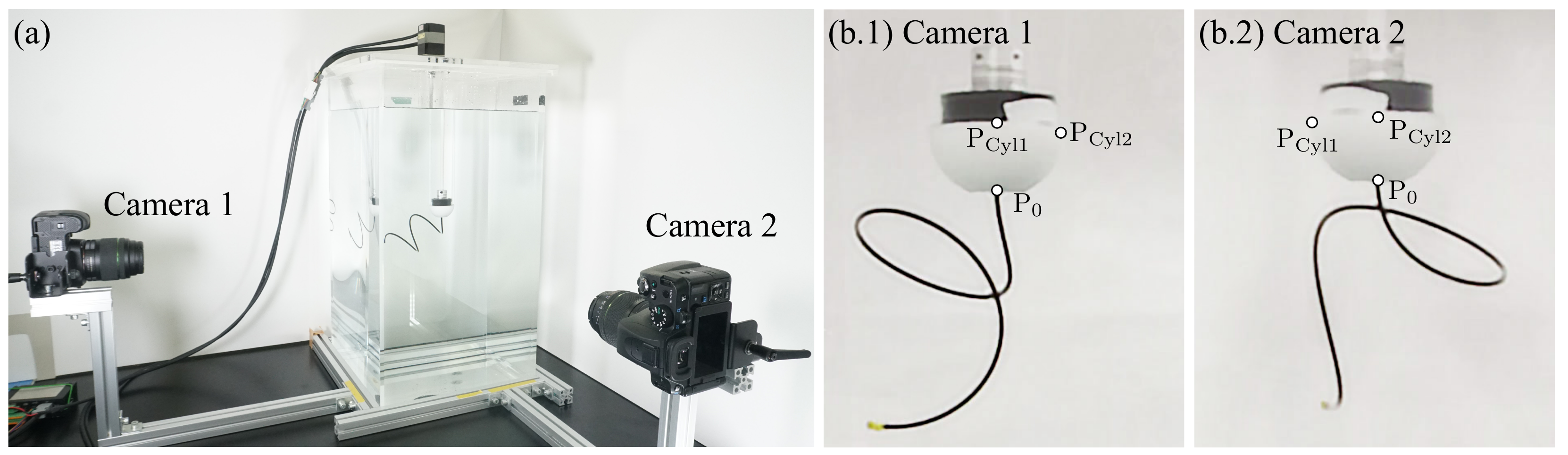}
    \caption{
    		Stereoscopic reconstruction of motion of a free end of a helical rod during wrapping.
		(a) Overview of our 3D tracking system:
		Two cameras positioned at right angles simultaneously capture the rod configuration.
		(b) Calibration procedure. (b.1) Image captured by Camera 1; (b.2) image captured by Camera 2.
		Here, P$_0$ is the top end of the helix, $\mathrm{P_{Cyl1}}$ is the closest point on the hemisphere as viewed from Camera 1, and $\mathrm{P_{Cyl2}}$ is the closest point on the hemisphere as viewed from Camera 2.
		Because the relative positions of P$_0$, $\mathrm{P_{Cyl1}}$ and $\mathrm{P_{Cyl2}}$ are known (the radius of the hemisphere is 2.5 mm) ,  the calibration of our method was conducted using these parameters.}
    \label{exp:fig_tracking:Exp}
\end{center}
\end{figure*}

\begin{figure*}[htbp]
\begin{center}
  \includegraphics[width=0.95\linewidth]{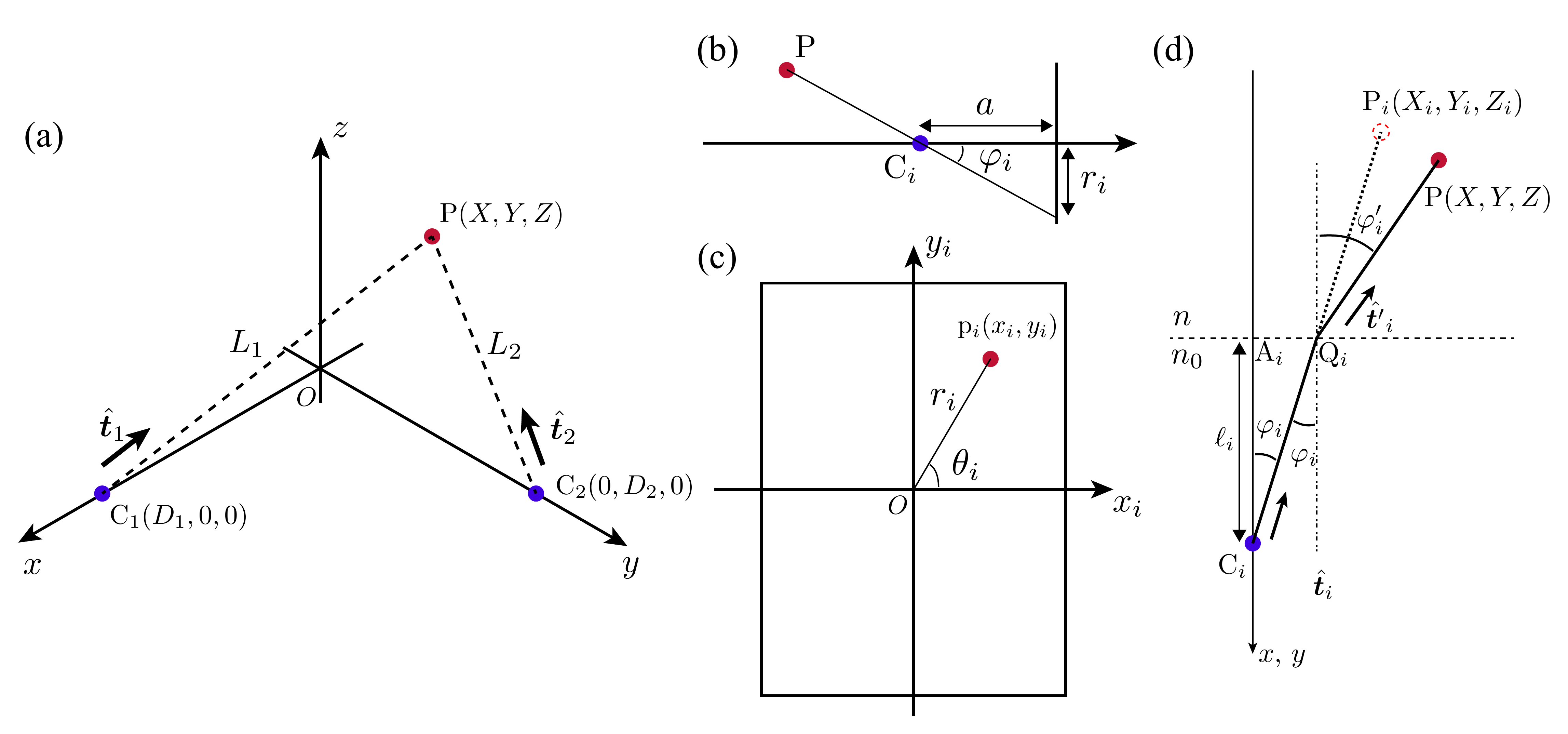}
    \caption{
		Description of our triangulation method.
		(a) Coordinate system and definition of the geometric parameters.
		(b) Distance $r$ in the picture space and apparent focal length $a$.
		(c) Coordinate system and configuration in the photographic image.
		(d) Coordinate system including the difference in the refractive indices between the media.
	}
    \label{exp:fig_tracking}
  \end{center}
\end{figure*}

The overview of the tracking system is shown in Fig.~\ref{exp:fig_tracking:Exp} (a).
Two digital cameras, denoted as C$_1$ and C$_2$ in Fig.~\ref{exp:fig_tracking:Exp}, with apparent focal length $a$ in the photographic space are placed at the same height with their optical axes set in the right angle. 
The coordinates of the tip of the rod (yellow) in the picture plane were obtained by processing the photograph images using ImageJ software. 
We then reconstructed its 3D coordinates in real space by applying a scheme similar to a standard triangulation method (detailed below). 
We have calibrated our method as described in Fig.~\ref{exp:fig_tracking:Exp} (b).

\subsection{Uniform refractive index case}
In the following, we describe our method of computing the tip position of rod P: $ (X, Y, Z)$ from the photographic images of C$_{1, 2}$.
In the Cartesian coordinate system $xyz$ defined in Fig.~\ref{exp:fig_tracking} (a), the positions of two cameras are given by 
C$_1 (D_1, 0, 0)$ and C$_2 (0, D_2, 0)$.
Writing $\overrightarrow{\mathrm{C}_i\mathrm{P}}=L_i\hat{\bm{t}_i}$, where $\hat{\bm{t}_i}=\overrightarrow{\mathrm{C}_i\mathrm{P}}/|\overrightarrow{\mathrm{C}_i\mathrm{P}}|$, we have $\overrightarrow{\mathrm{O}\mathrm{P}}=\overrightarrow{\mathrm{O}\mathrm{C}_i}+L_i\hat{\bm{t}}_i$.
Because $\overrightarrow{\mathrm{O}\mathrm{C}_i}$ is known, we can obtain $\overrightarrow{\mathrm{O}\mathrm{P}}$ by determining $\hat{\bm{t}_i}$ and $L_i$ from the images of C$_i$.

Hence, we first determine the polar and azimuthal angles $(\varphi_i,\theta_i)$ of $\hat{\bm {t}}_i$, where $i=1,2$.
Let $\varphi_1$ and $\varphi_2$ be the angles between $\hat{\bm{t}_1}$ and $-\bm{e}_x$ and $\hat{\bm{t}_2}$ and $-\bm{e}_y$, respectively.
Writing the position of P on the photographic image as p$_i (x_i, y_i)$, we can obtain $\varphi_i=\tan^{-1}{(a/r_i)}$ when $D_i \gg a$, where $r_i=\sqrt{x_i^2+y_i^2}$ and $a$ is the apparent focal length. 
In our measurement, $a=1680$ pixels. 
In addition, we have $\theta_i=\mathrm{sgn}{ (y_i)}\cos^{-1}{ (x_i/r_i)}$ for $i=1$ and 2, which results in 
$\hat{\bm{t}}_1=(-\cos{\varphi_1}, \sin{\varphi_1}\cos{\theta_i}, \sin{\varphi_1}\sin{\theta_1})$,
$\hat{\bm{t}}_2=(-\sin{\varphi_2}\cos{\theta_2}, -\cos{\varphi_2}, \sin{\varphi_2}\sin{\theta_2})$.

Expressing the components of $\overrightarrow{\mathrm{O}\mathrm{P}}$ in terms of the parameters related to C$_1$ and C$_2$, we have
\begin{align}
	X &= D_1 - L_1\cos{\varphi_1} 			= -L_2\sin{\varphi_2}\cos{\theta_2}, 	\label{eq:tracking:X} \\
	Y &= L_1\sin{\varphi_1}\cos{\theta_1}	=  D_2 - L_2\cos{\varphi_2}, 		 	\label{eq:tracking:Y} \\
	Z &= L_1\sin{\varphi_1}\sin{\theta_1}	=  L_2\sin{\varphi_2}\sin{\theta_2}. 	\label{eq:tracking:Z}
\end{align}
Solving these equations in terms of $L_1$ and $L_2$, we obtain
\begin{equation}
	L_1=\frac{D_1\tan{\theta_2}}{\cos{\varphi_1}\tan{\theta_2}-\sin{\varphi_1}\sin{\theta_1}},
	\quad
	L_2=\frac{D_2\tan{\theta_1}}{\cos{\varphi_2}\tan{\theta_1}+\sin{\varphi_2}\sin{\theta_2}}. 
 \label{eq:tracking:Li}
\end{equation}
When we obtain the numerical values of $L_{1,2}$ from Eq.~(\ref{eq:tracking:Li}), we can finally determine P$(X,Y,Z)$ from Eqs.~(\ref{eq:tracking:X})--(\ref{eq:tracking:Z}).

\subsection{Different refractive indices case}
The above formulation should be slightly modified to consider the refractive index of glycerin $n=1.47$ at 20$^{\circ}$, although the corrections of this to the final results become subtle.
Because the refractive index of the acrylic wall is 1.49 and is quite similar to that of glycerin, these are treated as the single background medium of the refractive index $n$.

Let the distance between the cameras be C$_i$ and the tank wall be $\ell_{i}$. 
These are the known distances [Fig.~\ref{exp:fig_tracking} (d)].
Snell's law suggests $n_0\sin{\varphi_i}=n\sin{\varphi^{\prime}_i}$, where $n_0$ is the refractive index of the air.
Subsequently, repeating the same argument presented above, we can finally obtain
\begin{eqnarray}
 \left(
\begin{array}{c}
  X \\
  Y \\
  Z    
\end{array}
\right)
 &=& 
\left(
\begin{array}{c}
 D_1 - \ell_1 - L_1\cos \varphi^{\prime}_1 \\
 \ell_1\tan{\varphi_1}\cos{\theta_1} + L_1\sin{\varphi^{\prime}_1}\cos{\theta_1} \\
 \ell_1\tan{\varphi_1}\sin{\theta_1} + L_1\sin{\varphi^{\prime}_1}\sin{\theta_1}
  \end{array}
\right), 
\label{eq:tracking:XYZ:1:n}
\end{eqnarray}
where the numerical values of $L_1$ and $L_2$ can be obtained from 
\begin{equation}
 L_1 = \frac{D_1\tan{\theta_2}+\ell_1 (\tan{\varphi_1}\sin{\theta_1}-\tan{\varphi_2})}{\cos{\varphi^{\prime}_1}\tan{\theta_2}-\sin{\varphi^{\prime}_1}\sin{\theta_1}}, 
 \quad
 L_2 = \frac{D_2\tan{\theta_1}-\ell_2 (\tan{\varphi_2}\sin{\theta_2}+\tan{\varphi_1})}{\cos{\varphi^{\prime}_2}\tan{\theta_1}+\sin{\varphi^{\prime}_2}\sin{\theta_2}}
 \label{eq:tracking:L1L2:n}
\end{equation}

\section{Post-buckling dynamics}

\begin{figure*}[h]
	\centering
	\includegraphics[width=0.95\linewidth]{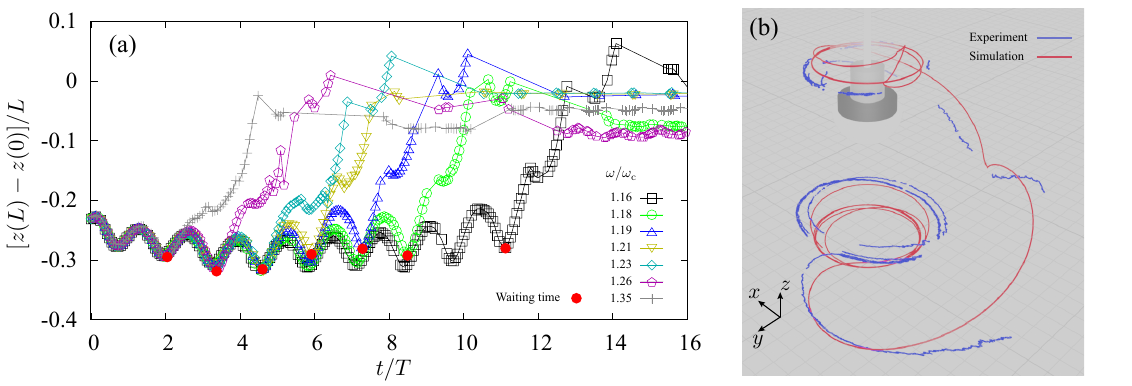}
	\caption{Slow buckling dynamics in the supercritical regime ($\omega/\omega_c>1$).
		(a) Rescaled vertical $(z)$ position of the free end of the rod, $[z(L)-z(0)]/L$, as a function of the number of revolutions $t/T$ from $t=0$, at which the motor starts rotating CW.
		A left-handed helical rod with pitch angle $\psi=70^{\circ}$, helical radius $R/R_{\mathrm{cell}}=0.6$, helical length $L=300$ mm, and helical turn $n=3.0$ is used for all cases.
		The helix was first rotated CCW for 10 turns at the frequency $f=0.040$ Hz ($t<0$) and then rotated CW at $f=0.058-0.120$ Hz from $t=0$.
		The red points indicate the waiting time $t_c$ for each $\omega/\omega_c$.
		(b) Reconstructed three-dimensional trajectory of the free end of the helical rod. 
		A helix $\psi=70^{\circ}$, $R/R_{\mathrm{cell}}=0.6$ and $n=2.0$ was used.
		The blue line shows the experimental data, and the red line shows the numerical simulation data for $\omega/\omega_{\mathrm{c}}=2.05$.  
	}
	\label{fig:buckling}
\end{figure*}

Figure~\ref{fig:buckling} (a) shows the time changes of the tip position of a helical rod, $z(L)$, during the wrapping instability for various angular velocities $\omega$.
As explained in the main text, the waiting time $t_c$ is defined as the time for the rod start moving upwards after the motor switches its rotation from CCW to CW. 
Thus, the points highlighted by the filled red symbols in Figure~\ref{fig:buckling} (a) specify the waiting times $t_c$ for different values of $\omega$ (indicated in the legend).
In Fig.~6 in the main paper, we adjusted $\omega_c$ to obtain the best scaling behavior; $\omega_c$s in Fig.~6 are assumed 1.115 times larger than those shown in the lengend in Fig.~\ref{fig:buckling} (a).
This level of the adjustment is acceptable considering the intrinsic ambiguity of the prefactor 0.06 in the scaling relation for $\omega_c$ determined in the phase diagram in Fig.~5.

\section{Parameter values of flagellar-wrapping bacteria}
Here, we detail the numerical values for the three bacterial data points on the phase diagram (Fig.~5) in the main text, based on the corresponding references.
First, the Young's modulus $E$ and radius of cross section $a$ of a flagellar filament are set as $E=4.5\times 10^8~\mathrm{Pa}$ and $a=10~\mathrm{nm}$.

A polar flagellated bacterium typically has a few flagella at one pole that form a bundle during propulsion. 
In this study, we assume the bundle of flagella as a single elastic helical filament. 
Assuming $B_n$ as the bending stiffness of a bundle of $n$ flagella, we expect $B_n=nB_1$, where $B_1=\pi Ea^4/4$.

\subsubsection{{\it Caballeronia insecticola} (synonym {\it Burkholderia} sp. RPE64): Ref.~\cite{Kinosita-ISME-2018}}
The measured values of the helix radius and pitch are summarized in Supplemental Table 1 of Ref.~\cite{Kinosita-ISME-2018}. 
We calculate the arclength of a flagellar filament, $L$, from $L=L_{ee}/\cos\psi$, where we interpret the end-to-end distance of the filament $L_{ee}=6$ $\mu$m from the image given in the main paper~\cite{Kinosita-ISME-2018}, with the pitch angle of the Normal form $\psi=\tan^{-1}(2\pi R/P)=32^{\circ}$.  
While the resulting value $L=7.1$ $\mu$m is slightly higher than the average value $5.6$ $\mu$m reported in Supplemental Table 1~\cite{Kinosita-ISME-2018}, we use it as the representative value of $L$ for our analysis, because it gives a 1.5--2 turn helix for the coil form, as typically observed.
We obtain the rotational frequency of $f=40$ Hz observed in the $0.3\%$ methylcellulose solution, the viscosity of which is 3.5 cps.

\subsubsection{{\it Shewanella putrefaciens} (CN32): Ref~\cite{Kuhn-PNAS-2017}}
The parameters of {\it S. putrefaciens} are summarized in Table S4 in Supplemental Information in Ref.~\cite{Kuhn-PNAS-2017}.
Specifically, we used $R=0.473~\mathrm{\mu m}$, $P=1.44~\mathrm{\mu m}$, and $L=6.5~\mathrm{\mu m}$.
Because {\it S. putrefaciens} is a bacterium with a primary single polar flagellum, we assumed $n=1$.
The viscosity of the surrounding fluid is in the range of $\eta=2-33~\mathrm{mPa\cdot s}$.
We used $\eta=5~\mathrm{m Pa\cdot s}$ for our plot.
The rotational frequency of the filament $f$ [Hz] is not given explicitly.
From Movie S2\cite{Kuhn-PNAS-2017}, the period of a full single turn of the coil was determined to be approximately 20 ms, from which we estimated $f=50~\mathrm{Hz}$.

\subsubsection{{\it Pseudomonas putida}: Ref~\cite{Hintsche-SciRep-2017}}
The parameters of {\it P. putida} are described in the paragraph "Swimming with the flagella wrapped around the cell body", in the main text of Ref.~\cite{Hintsche-SciRep-2017}.
Specifically, we used $R=0.6~\mathrm{\mu m}$ and $P=2.1~\mathrm{\mu m}$ for the coil configuration.
Although the contour length of the filament is not given explicitly, we estimated it as $L=8.0~\mathrm{\mu m}$ because we obtained the end-to-end distance of $\sim 5$ $\mu$m with the pitch angle for the normal form as $\psi=51^{\circ}$.
The number of flagellar filaments is mentioned as "multiple" in the main text but no specific values are given~\cite{Hintsche-SciRep-2017}, so we assumed $n=3$.
We assumed the rotational frequency as 50 Hz at wrapping.
Because the viscosity of the surrounding liquid was not described particularly in Ref.~\cite{Hintsche-SciRep-2017}, we assumed $\eta=0.2~\mathrm{mPa\cdot s}$.

\section*{Supplemental Movies}
The description of the relevant parameters -- $L$: arclength of a helix, $R$: radius of a helix, $P$: pitch of a helix, $n$: the number of helical turns, $\psi$: pitch angle of a helix, $f$: CW rotational frequency of a stepping motor, $\omega=2\pi\times f$: angular velocity of a helix at the clamped end, $\omega_c$: the critical angular velocity of the buckling of a helix.  \\

\noindent
{\bf SI Movie 1:}
Cell behavior of wild type {\it C. insecticola} captured with phase-contrast microscopy. Area $300 \mu$m $\times$ $300 \mu$m. \\

\noindent
{\bf SI Movie 2:}
Twirling helix (Coil form) observed in the scaled experiment.  $L=358$ mm, $R=22.5$ mm, $P=57.1$ mm, $n=2.35$, $\psi=68^{\circ}$,
$f=0.01$ Hz, $\omega/\omega_c=0.86$. Video replay at 100x speed. \\

\noindent
{\bf SI Movie 3:}
Wrapping helix (Coil form) observed in the scaled experiment.  $L=400$ mm, $R=30.0$ mm, $P=68.0$ mm, $n=2$, $\psi=70^{\circ}$,
$f=0.006$ Hz, $\omega/\omega_c=1.97$. Video replay at 111x speed. \\

\noindent
{\bf SI Movie 4:}
Wrapping helix (Coil form) observed in the numerical simulation with the long-ranged hydrodynamic interactions. 
$N=61$, $n=1.93$, $\psi=71^{\circ}$, $\omega/\omega_c=2.05$. \\

\noindent
{\bf SI Movie 5:}
Folding helix (Coil form) observed in the scaled experiment.  $L=358$ mm, $R=22.5$ mm, $P=57.1$ mm, $n=2.35$, $\psi=68^{\circ}$,
$f=0.02$ Hz, $\omega/\omega_c=1.72$. Video replay at 50x speed.\\

\noindent
{\bf SI Movie 6:}
Folding helix (Normal form) observed in the scaled experiment.  $L=359$ mm, $R=12.5$ mm, $P=120$ mm, $n=2.5$, $\psi=33^{\circ}$,
$f=0.10$ Hz, $\omega/\omega_c=1.29$. Video replay at 10x speed.\\

\noindent
{\bf SI Movie 7:}
Critical dynamics of the buckling (Coil form) observed in the scaled experiment.  $L=300$ mm, $R=15$ mm, $P=34$ mm, $n=3.0$, $\psi=70^{\circ}$,
$f=0.10$ Hz, $\omega/\omega_c=1.04$. Video replay at 17x speed.\\

\noindent
{\bf SI Movie 8:}
Critical dynamics of the buckling (Coil form) observed in the numerical simulation with the long-ranged hydrodynamic interactions. 
$N=61$, $n=3$, $\psi=70^{\circ}$, $\omega/\omega_c=1.05$. 

\end{document}